\documentclass[12pt]{article}
\usepackage{amsmath}
\usepackage{amssymb}
\usepackage{graphicx}
\usepackage{float}
\usepackage{indentfirst}
\usepackage{mathrsfs}
\usepackage{dsfont}%
\usepackage{multirow}
\usepackage{diagbox}

\usepackage{setspace}

\usepackage[labelfont=bf,labelsep=period]{caption}

\usepackage[numbers,sort&compress]{natbib}

\newtheorem{theorem}{Theorem}

\newtheorem{observation}{Observation}

\title{Characterizing the Quantum Non-locality by the Mathematics of Magic Square \\ [0.7cm]}

\author
{Jun-Li Li and Cong-Feng Qiao$^{\ast}$ \\ [0.2cm]
\normalsize{School of Physical Sciences, University of Chinese Academy of Sciences,} \\
\normalsize{YuQuan Road 19A, Beijing 100049, China}\\ [2pt] \normalsize{Center of Materials Science and Optoelectronics Engineering \& CMSOT,} \\
\normalsize{University of Chinese Academy of Sciences, YuQuan Road 19A, Beijing 100049, China}\\ [2pt]
\normalsize{Key Laboratory of Vacuum Physics, University of Chinese Academy of Sciences} \\
\normalsize{YuQuan Road 19A, Beijing 100049, China} \\ [3mm]
\normalsize{$^\ast$ To whom correspondence should be addressed; E-mail: qiaocf@ucas.ac.cn.}
}

\date{}

\begin{document}
\baselineskip24pt \maketitle
\begin{abstract} \doublespacing
By constructing the quantum state in high-dimensional probability tensor, we find the quantum magic square(QMS) may stand as an ideal means of characterizing the non-local phenomena, i.e. the separability, entanglement, two/one-way steering, and Bell non-locality, etc. In this scheme, different types of non-locality exhibit distinctive inner structures of the probability tensor, which are observable in form of the partial sum of the tensor components. In application, we prove the Bell and GHZ theorems, and demonstrate that the uncertainty relation may rate the non-locality, from Bell locality to separability. We derive a conditional majorization uncertainty relation, which is superior to the steering criterion previously thought to be optimal for the uncertainty relation.
\end{abstract}

\newpage

\section{Introduction}

Entanglement is a unique nature of the quantum world, which exhibits in certain ways not manifesting in classical physics, for instance the non-locality, and plays a key role in implementing quantum information tasks. The advent and first application of the entangled state may date back to the EPR paradox \cite{EPR-Paradox}, by which Einstein, Podolsky, and Rosen questioned the completeness of the quantum mechanics by means of the state with two correlated systems. Schr\"odinger coined the term ``entanglement'' to describe the peculiar connection in such correlated systems: One may steer part of the system in spite of no access to it \cite{Ent-Stee-Schrodinger}. To exhibit the non-locality of the entangled system, Bell inequality was first introduced as early as the year 1964 \cite{Bell-1964}, while the quantum steering did not stand as a distinct nonlocal phenomenon from Bell non-locality until 2007 \cite{Steer-1}.

Bell non-locality can be understood as the violation of various forms of Bell inequalities \cite{CHSH,CGLMP}, whereas, the ascertainment of quantum steering and entanglement is subject to different criteria \cite{Quantum-Steering, Separable-QL}, viz, the correlation function \cite{Separability-Bell,Steering-CHSH}, uncertainty relation \cite{Separability-Uncertainty, Steering-uncertainty}. Recently, some delicate measures \cite{Sep-cov, Steering-Moment} are employed to witness the entanglement or steerability. Up to now, finding the practical necessary and sufficient conditions remains to be an open question for both steering and entanglement. The criteria based on the uncertainty relation usually have distinct motivation and better performance \cite{Entropy-steering}. However, the optimal lower bound of the uncertainty relation, which is crucial in detecting quantum steering or separability, turns out to be another challenging task \cite{Entropy-app}. Very recently, the optimal bound problem is solved for the universal uncertainty relation by virtue of the lattice theory \cite{Maj-opt}, which opens a new horizon for the characterization of different nonlocal phenomena based on uncertainty relation.

In this paper, we propose a quantum magic square(QMS) scheme to describe the quantum non-locality, in which the local randomness of different observables are incorporated into a high-dimensional probability tensor with given marginal distributions. With this scenario, the Bell and GHZ theorems are explicitly and uniformly reexhibited and the in fact the manipulation can be smoothly extended to the arbitrary multipartite system. The hierarchical structure of entanglement, from the Bell local state to the separable one, emerges in the process of successive application of the uncertainty relation. As an example, the steering criterion applicable to any number of observables is given based on the optimal majorization uncertainty relation, which found is superior to what thought to be optimal before.

\section{The magic square and quantum non-locality}

\begin{figure}[htbp]\centering
\scalebox{0.55}{\includegraphics{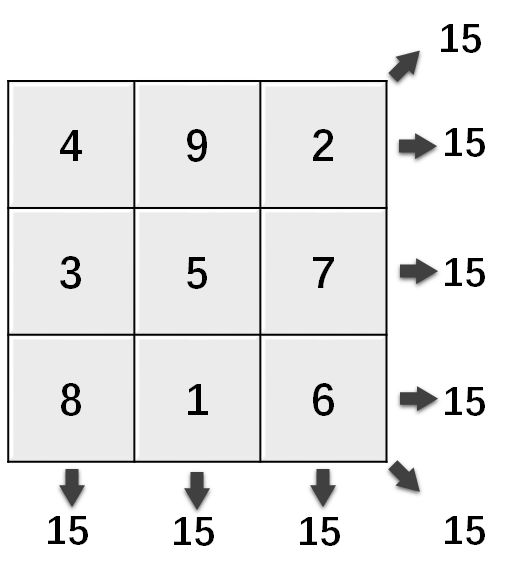}}\hspace{-0.8cm}
\caption{The $3\times 3$ magic square. In classical magic square, the row, column, and diagonal all sum to 15.} \label{33Luoshu}
\end{figure}

A classical magic square is an $n\times n$ square grid filled with distinct positive integers such that each cell contains a different integer and the sum of the integers in each row, column and diagonal satisfies given constraints, i.e., the sums are all equal, see Figure \ref{33Luoshu}. In a bipartite system, the joint measurements $X$ and $Y$ on each particle lead to a joint distribution $P(X,Y)$. And, two different distributions $P(X,Y)$ and $P(X',Y')$ may be regarded as the marginal distributions of high dimensional distribution $m_{X,X',Y,Y'}$. Noticing of this, we can then treat $m_{X,X',Y,Y'}$ as a quantum magic square with marginals $P(X,Y)$ and $P(X',Y')$. Following, we show how the different types of non-locality emerge in filling the $m_{X,X',Y,Y'}$ elements.

\subsection{Magic square for quantum entanglement}

\begin{figure}[htbp]\centering
\scalebox{0.47}{\includegraphics{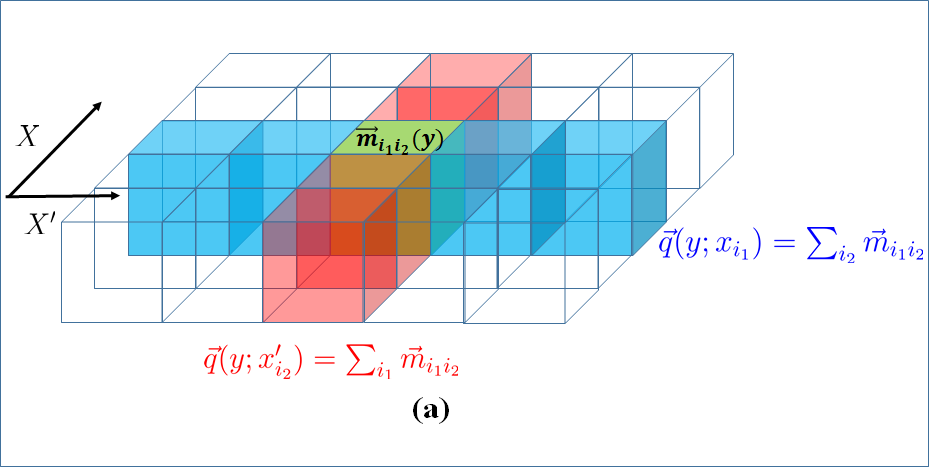}}
\scalebox{0.47}{\includegraphics{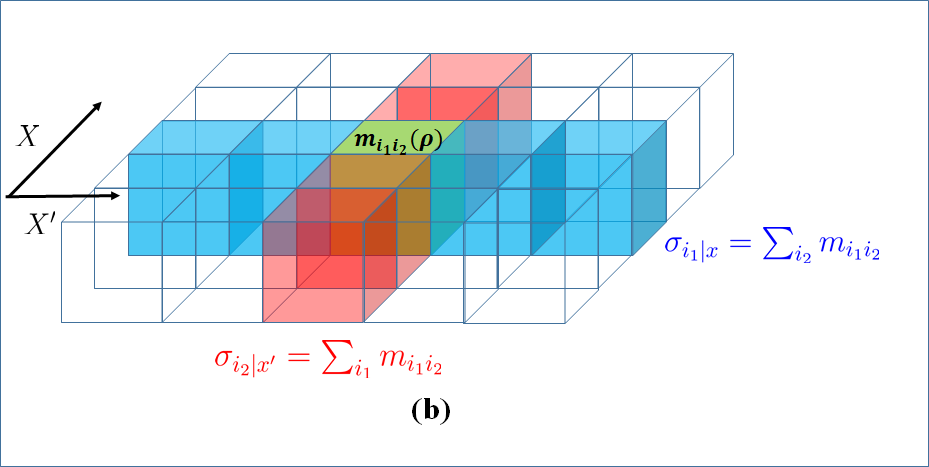}}
\caption{The magic square representation of bipartite system. {\bf (a)} The cubic cells for $\vec{m}_{i_1i_2}(y)$, the unnormalized distribution vectors, and $\vec{q}(y;x_{i_1})=\sum_{i_2}\vec{m}_{i_1i_2}(y)$; {\bf (b)} The cubic cells for $m_{i_1i_2}(\rho)$, the unnormalized state, and $\sigma_{i_1|x} = \sum_{i_2} m_{i_1i_2}(\rho)$ is the assemblage by summing over $i_2$. Similarly, the summation over column yields $\vec{q}(y;x_{i_2}')$ and $\sigma_{i_2|x'}$.} \label{Figure-S-1}
\end{figure}

With quantum state $\rho_{AB}$ of a bipartite system $A(lice)$ and $B(ob)$, we may measure certain observables $X$ and $Y$ on each side and obtain the joint distribution $P(X,Y)$. If $\rho_{AB}$ is entangled and exhibits Bell non-locality, the joint distribution can not admit the following decomposition
\begin{align}
\forall \, X\, \mathrm{and}\, Y \; , \; P_{ij}(X,Y) & = \sum_{\lambda} \kappa_\lambda \cdot [p^{(\lambda)}_{i}(x) q^{(\lambda)}_{j}(y)] \; . \label{decom-steer-xy}
\end{align}
Here $p_{i}^{(\lambda)}(x)$ and $q_{j}^{(\lambda)}(y)$ are normalized distributions of the measurements on $X$ and $Y$ with outcomes $x_{i}$ and $y_{j}$, respectively. $\lambda$ denotes the possible hidden variable involving in the measurement, and $\kappa_\lambda$ is the normalized weight of quanta source. For different measurements $X$ and $X'$ on $A$, the joint distributions $P(X,Y)$ and $P(X',Y)$ can be expressed as
\begin{align}
P_{i_1 j}(X,Y) = \sum_{i_2} m_{i_1i_2 j} \; , \;
P_{i_2 j}(X',Y) = \sum_{i_1} m_{i_1i_2 j}  \; . \label{Decom-zx-y}
\end{align}
where $m_{i_1i_2 j} := \sum_\lambda \kappa_\lambda \cdot [ p^{(\lambda)}_{i_1}(x) p^{(\lambda)}_{i_2}(x')] q^{(\lambda)}_j(y)$ constitutes the components of the unnormalized distribution $\vec{m}_{i_1i_2}(y)$. The probability distribution vectors on the observation of $Y$ conditioned on the measurement results of $x_{i_1}$ or $x'_{i_2}$ can be defined as
\begin{align}
\vec{q}(y|x_{i_1}) := \frac{\vec{q}(y;x_{i_1})}{p(x_{i_1})}  \; , \; \vec{q}(y|x'_{i_2})  := \frac{\vec{q}(y;x'_{i_2})}{p(x'_{i_2})} \;. \label{cell-sum}
\end{align}
Here $\vec{q}(y;x_{i_1}) := \sum_{i_2} \vec{m}_{i_1i_2}(y)$ is the $i_1$th row of $P(X,Y)$ and $p(x_{i_1}) \equiv \sum_k q_k(y;x_{i_1})$. Similar definition applies to $\vec{q}(y;x'_{i_2})$ as well, see Figure \ref{Figure-S-1}(a).

Furthermore, applying measurements $Y$ and $Y'$ on the $B$ side we can then get a tensor
\begin{align}
m_{i_1i_2j_1j_2} = \sum_{\lambda} \kappa_{\lambda} \cdot \left[p^{(\lambda)}_{i_1}(x) p^{(\lambda)}_{i_2}(x')\right] \left[q^{(\lambda)}_{j_1}(y) q^{(\lambda)}_{j_2}(y')\right] \; , \label{Corr-Calc}
\end{align}
the quantum magic square of bipartite state. Note, all observable distributions can be obtained from $m_{i_1i_2j_1j_2}$ by partial sums, i.e., $P(x)$, $P(y')$, $P(x,y')$, etc. The generalization to tripartite system with measurements $X$ and $X'$, $Y$ and $Y'$, $Z$ and $Z'$ on each side is straightforward,
\begin{align}
m_{i_1i_2j_1j_2k_1k_2} = \sum_{\lambda} \kappa_{\lambda} \cdot \left[p^{(\lambda)}_{i_1}(x) p^{(\lambda)}_{i_2}(x')\right] \left[q^{(\lambda)}_{j_1}(y) q^{(\lambda)}_{j_2}(y')\right] \left[r^{(\lambda)}_{k_1}(z) r^{(\lambda)}_{k_2}(z')\right] \; . \label{Sudoku-tensor}
\end{align}
Here $m_{i_1i_2j_1j_2k_1k_2}$ is the quantum magic square of tripartite state, and it is easy to verify that
\begin{align}
m_{i_1i_2j_1j_2k_1k_2}\geq 0\; , \; \sum_{i_1,i_2,j_1,j_2,k_1,k_2} m_{i_1i_2j_1j_2k_1k_2} = 1\;. \label{Sudoku-tensor-condition}
\end{align}

From the definition of probability tensor, the quantum magic square, one may have the following observation
\begin{observation}
If the system is Bell local, there will be a quantum magic square representation for the quantum state, and vice versa.
\end{observation}
The observation sets up an equivalence relation for Bell locality and magic square representation for quantum state. Hence, the non-existence of magic square representation amounted to the Bell non-locality of quantum state.

According to Wiseman, {\it et al}. \cite{Steer-1}, if $\rho_{AB}$ is entangled but $A$ cannot steer the system of $B$, then $B$ admits the following decomposition
\begin{align}
\forall X\;, \; \sigma_{i|x} = \sum_{\lambda} \kappa_{\lambda} \cdot p^{(\lambda)}_{i}(x) \rho^{(\lambda)}\; . \label{decom-rho}
\end{align}
Here, $\sigma_{i|x}$ named assemblage describes the unnormalized quantum state of $B$ conditioned with the observation of $x_{i}$ on $A$ side \cite{Assem-d}. The magic square representation for quantum steering is obtainable by inserting a normalized distribution into equation (\ref{decom-rho}) as follows. For two measurements $X$ and $X'$ on $Alice$, we are allowed to define the following assemblages
\begin{align}
\sigma_{i_1|x} := \sum_{i_2} m_{i_1i_2}(\rho) \; , \;
\sigma_{i_2|x'} := \sum_{i_1} m_{i_1i_2}(\rho) \; , \label{Assemblage-xx}
\end{align}
where $m_{i_1i_2}(\rho) = \sum_{\lambda} \kappa_{\lambda}\ [p^{(\lambda)}_{i_1}(x) p^{(\lambda)}_{i_2}(x')] \rho^{(\lambda)}$ is an unnormlized quantum state and $p_{i_2}^{(\lambda)}(x')$ is the inserted normalized distribution with $\sum_{i_2} p_{i_2}^{(\lambda)}(x') = 1$, as shown in Figure \ref{Figure-S-1}(b). Evidently, extending to the multiple measurements on one side, one may readily obtain $m_{i_1\cdots i_M}(\rho)$, and hence the Observation:
\begin{observation}
If the system is Bell local and $A$ cannot steer the system of $B$, then there exists the quantum magic square representation $m_{i_1\cdots i_M}(\rho)$ of the quantum state for arbitrary measurements $X^{(1)},\cdots, X^{(M)}$ on $A$, and vice versa.
\end{observation}
From this Observation, it is transparent to formulate the steerability of $B$ to $A$. Moreover, the observation sets up an equivalent relation between quantum steering and the magic square representation of quantum state within the Bell local system. Following we apply the QMS representation to some typical non-local phenomena to show its capacity in characterizing the quantum non-locality.

\subsection{Bell non-locality in QMS}

\begin{figure}\centering
\scalebox{0.6}{\includegraphics{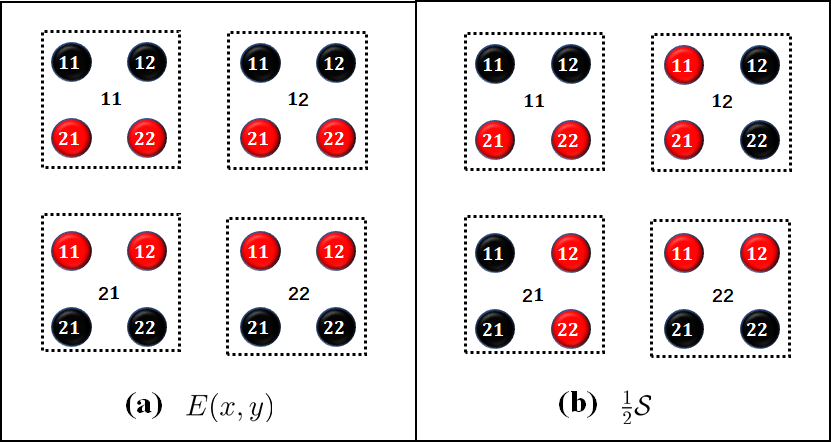}}
\caption{The magic squares for CHSH inequality of qubit system. Each circled node represents a tensor element $m_{i_1i_2j_1j_2}$. The numbers centered in the dashed square are $i_1$ and $i_2$ and the numbers centered in the circled nodes are $j_1$ and $j_2$. {\bf (a)} The correlation function, $E(x,y) = \sum_{\{\mathrm{black}\}} m_{i_1i_2j_1j_2} - \sum_{\{\mathrm{red}\}} m_{i_1i_2j_1j_2}$; {\bf (b)} The sum of four correlation functions, $\mathcal{S}=E(x,y) - E(x,y') + E(x',y) + E(x',y')$.} \label{Fig-CHSH}
\end{figure}

It is well-known that the CHSH inequality for qubit system writes \cite{CHSH}
\begin{align}
\mathcal{S} = E(x,y) - E(x,y') + E(x',y) + E(x',y') \in [-2,2] \; , \label{CHSH-ineq}
\end{align}
where $E(\cdot,\cdot)$ denotes the correlation function of measurement $X$ and $X'$ on $Alice$ and $Y$ and $Y'$ on $Bob$. Considering $m_{i_1i_2j_1j_2}$ in equation (\ref{Corr-Calc}) and taking $E(x,y)$ as an example, we have
\begin{align}
E(x,y) = P_{11}(x,y)- P_{12}(x,y)- P_{21}(x,y)+ P_{22}(x,y) \; ,
\end{align}
where $P_{i_1j_1}(x,y) = \sum_{i_2,j_2} m_{i_1i_2j_1j_2}$. Then, from the magic square in Figure \ref{Fig-CHSH}(a) we have
\begin{align}
E(x,y) = \sum_{\{\mathrm{black}\}} m_{i_1i_2j_1j_2} - \sum_{\{\mathrm{red}\}} m_{i_1i_2j_1j_2} \; , \label{Exy-Sudoku}
\end{align}
and similarly other three correlation functions. The sum of four correlation functions are found to be $\mathcal{S} = 2\left( \sum_{\{\mathrm{black}\}}  m_{i_1i_2j_1j_2}- \sum_{\{\mathrm{red}\}} m_{i_1i_2j_1j_2} \right)$. The $\frac{1}{2}\mathcal{S}$ is intuitively illustrated in Figure \ref{Fig-CHSH}(b), from which one can easily read
\begin{align}
|\mathcal{S}| = \left| E(x,y) - E(x,y') + E(x',y) + E(x',y') \right|  \leq 2 \; . \label{CHSH-squares}
\end{align}
The proof of equation (\ref{CHSH-squares}) is given in the Appendix.

\begin{figure}\centering
\scalebox{0.6}{\includegraphics{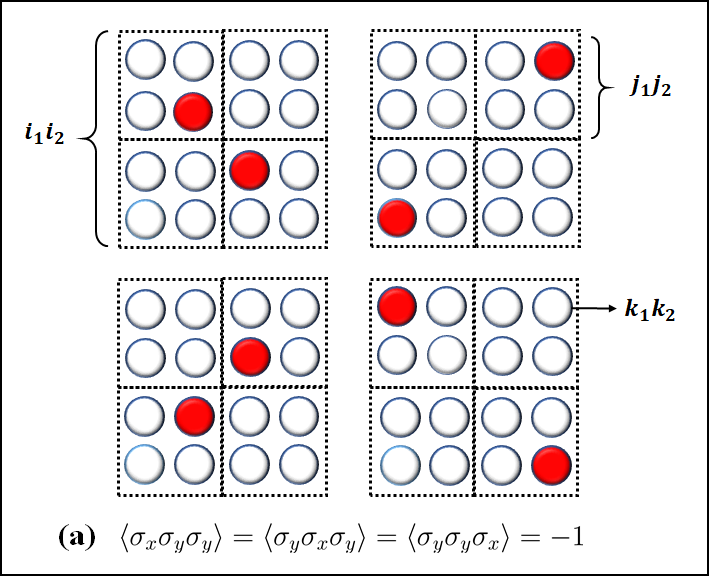}}
\scalebox{0.6}{\includegraphics{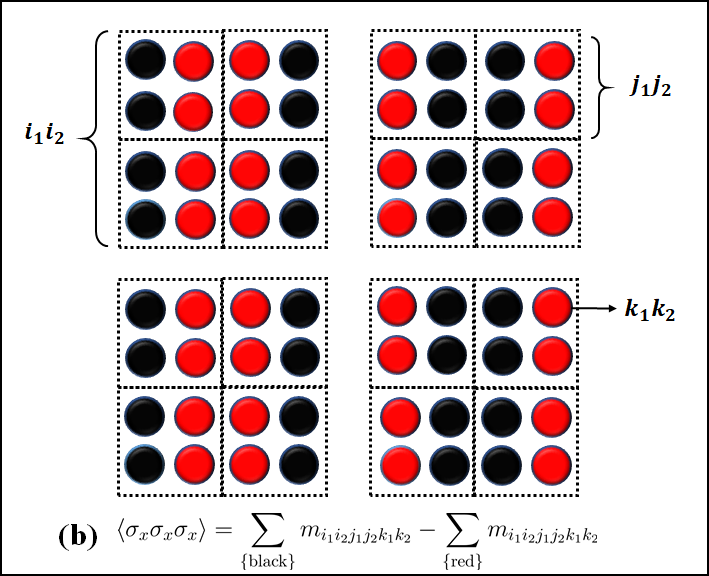}}
\caption{The magic squares for GHZ theorem. {\bf (a)} The circled nodes $m_{i_1i_2j_1j_2k_1k_2}$, which and only which satisfying $\langle \sigma_x \sigma_{y} \sigma_y \rangle = \langle \sigma_y \sigma_{x} \sigma_y\rangle = \langle \sigma_y\sigma_{y} \sigma_x\rangle = - 1$, yields $\sum_{\{\mathrm{red}\}} m_{i_1i_2j_1j_2k_1k_2} =1$ and $\sum_{\{\mathrm{white}\}} m_{i_1i_2j_1j_2k_1k_2} =0$; {\bf (b)} The black nodes $m_{i_1i_2j_1j_2k_1k_2}$ result in $\langle \sigma_x\sigma_x\sigma_x \rangle = 1$ and the red ones lead to $\langle \sigma_x\sigma_x\sigma_x \rangle = -1$. } \label{Fig-GHZ}
\end{figure}

By means of QMS the demonstration of GHZ theorem is even simpler. Consider the GHZ state $|\psi\rangle = \frac{1}{2}(|+++\rangle + |---\rangle)$, it is the eigenvector of four joint osbervables $\sigma_x \sigma_{y} \sigma_y$, $\sigma_y \sigma_{x} \sigma_y$, $\sigma_y\sigma_{y} \sigma_x$, and $\sigma_x\sigma_{x} \sigma_x$ with eigenstates of $-1$, $-1$, $-1$, and $+1$ respectively. Given the tensor elements $m_{i_1i_2j_1j_2k_1k_2}$ in equation (\ref{Sudoku-tensor}), the expectation values $\langle \sigma_x \sigma_{y} \sigma_y \rangle = \langle \sigma_y \sigma_{x} \sigma_y\rangle = \langle \sigma_y\sigma_{y} \sigma_x\rangle = - 1$ requires $\sum_{\{\mathrm{red}\}}m_{i_1i_2j_1j_2k_1k_2} =1$ and $\sum_{\{\mathrm{white}\}}m_{i_1i_2j_1j_2k_1k_2} =0$ according to Figure \ref{Fig-GHZ}(a) (derivation details are shown in Appendix). However, the expectation value $\langle \sigma_x\sigma_x\sigma_x\rangle$ in Figure \ref{Fig-GHZ}(b) tells
\begin{align}
\langle \sigma_x \sigma_x\sigma_x\rangle = \sum_{\{\mathrm{black}\}} m_{i_1i_2j_1j_2k_1k_2} - \sum_{\{\mathrm{red}\}} m_{i_1i_2j_1j_2k_1k_2} \; ,
\end{align}
which results in a contradiction, i.e., the red elements in Figure \ref{Fig-GHZ}(a) unavoidably leads to $\langle \sigma_x\sigma_x\sigma_x\rangle = -1$ in Figure \ref{Fig-GHZ}(b), against the quantum mechanics prediction $\langle \sigma_x\sigma_x\sigma_x\rangle = 1$.

In above we select for simplicity the bipartite and tripartite cases as examples to show how quantum magic square works in vindicating the Bell type inequalities. In practice, one can easily move forward to derive new Bell inequalities by QMS for high dimensional and multipartite system with arbitrary number of measurements \cite{CGLMP}.

\subsection{Quantum steering in QMS}

\begin{figure}[htbp]\centering
\scalebox{0.6}{\includegraphics{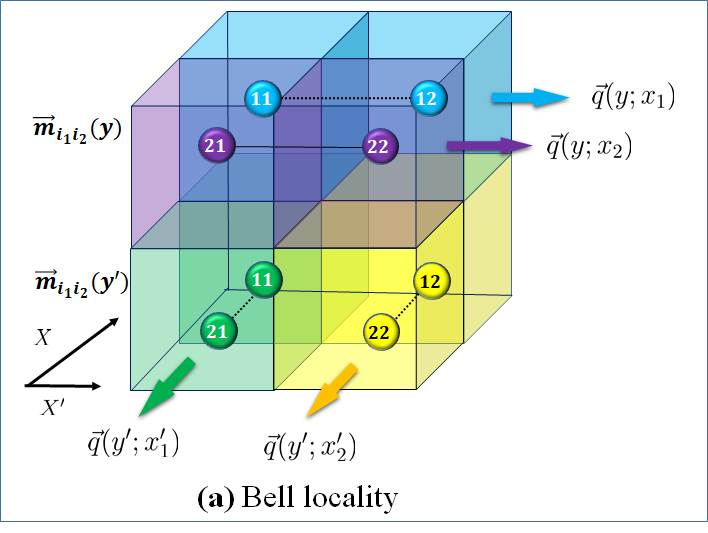}}
\scalebox{0.6}{\includegraphics{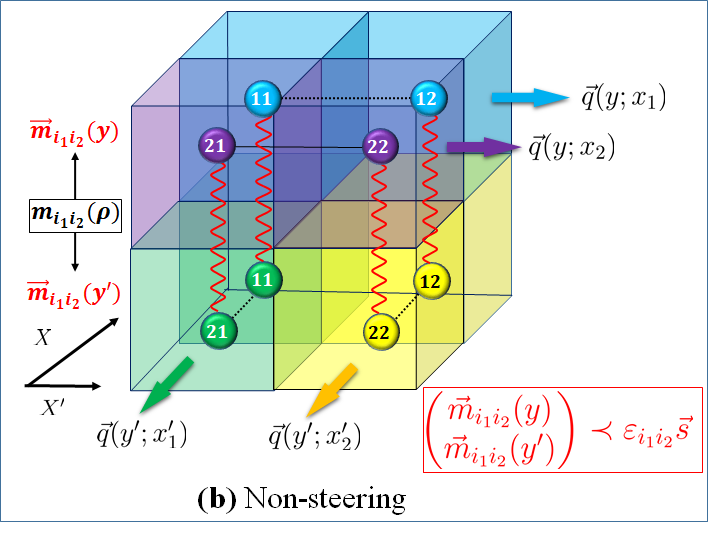}}
\caption{The Bell locality and non-steering are distinguished by magic squares. The circled nodes in upper and lower surfaces stand for $\vec{m}_{i_1i_2}(y)$ and $\vec{m}_{i_1i_2}(y')$ in observing $Y$ and $Y'$ respectively. {\bf (a)} $\vec{m}_{i_1i_2}(y)$ and $\vec{m}_{i_1i_2}(y')$ are independent distribution vectors for Bell locality; {\bf (b)} $\vec{m}_{i_1i_2}(y)$ and $\vec{m}_{i_1i_2}(y')$ are subject to the uncertainty relation in state $m_{i_1i_2}(\rho)$ when it is non-steerable from $A$ to $B$.} \label{Figure-2-S}
\end{figure}

In the framework of QMS, the difference between Bell locality and non-steerability is manifested in Figure \ref{Figure-2-S}, where without loss of generality we take the bipartite qubit system as an example. For Bell local state, $\vec{m}_{i_1i_2}(y)$ and $\vec{m}_{i_1i_2}(y')$ are two independent distributions due to the fact that $q^{(\lambda)}_{j_1}(y)$ and $q^{(\lambda)}_{j_2}(y')$ have no prior correlation in equation (\ref{Corr-Calc}), see Figure \ref{Figure-2-S}(a). However, for the state non-steerable from $A$ to $B$, $\vec{m}_{i_1i_2}(y)$ and $\vec{m}_{i_1i_2}(y')$ are constrained by the uncertainty relation within in the magic squares because they come from the same state $m_{i_1i_2}(\rho)$,
see Figure \ref{Figure-2-S}(b). From Ref. \cite{Maj-opt}, the majorization uncertainty relation
\begin{align}
\forall i_1,i_2\; , \;\vec{m}_{i_1i_2}(y) \oplus \vec{m}_{i_1i_2}(y') \prec \varepsilon_{i_1i_2} \vec{s}(y,y') \;  \label{Maj-cell}
\end{align}
holds, in which $\varepsilon_{i_1i_2} = \mathrm{Tr}[m_{i_1i_2}(\rho)]$ and $\vec{s}(y,y') $ is an optimal bound relying merely on observables $Y$ and $Y'$.

In $N$-dimensional quantum system, for $M$ different measurements $X^{(i)}$ on $Alice$, following theorem exists:
\begin{theorem}
If $A$ cannot steer $B$, the conditional majorization uncertainty relation
\begin{align}
\bigoplus_{i=1}^{M} \left[\sum_{j=1}^N \vec{q}^{\,\downarrow}(y^{(i)}|x^{(i)}_{j})p(x_j^{(i)})\right] \prec    \vec{s}\; \label{Theorem-equation}
\end{align}
should be satisfied. Here $\vec{q}^{\,\downarrow}(y^{(i)}|x^{(i)}_{j})$ is the distribution of measuring $Y^{(i)}$ on $Bob$ conditioned on the measurement results of $x_{j}^{(i)}$ on $Alice$ with components being rearranged in descending order; $\vec{s}$ is the least upper bound for the majorization uncertainty relation of joint measurements $Y^{(1)}, \cdots, Y^{(M)}$. \label{Theorem-steering}
\end{theorem}
The demonstration of the Theorem \ref{Theorem-steering} is presented in the Appendix. To illustrate and verify its effectiveness, we apply it to the Werner and isotropic states as examples.

Two-dimensional Werner and isotropic states are equivalent, can take the following form
\begin{align}
\rho_{\mathrm{W}} = \frac{1-\eta}{4} \mathds{1} \otimes \mathds{1} + \eta|\psi^-_{12}\rangle \langle \psi^-_{12}| \; ,
\end{align}
where $\eta$ is a premeter, $|\psi^-_{12}\rangle = \frac{1}{\sqrt{2}}(|12\rangle - |21\rangle)$. From Theorem \ref{Theorem-steering}, for two-dimensional case, the joint measurements $X$ and $Y$ for $\sigma_x$, $X'$ and $Y'$ for $\sigma_y$ on Alice and Bob respectively give $\eta \leq 1/\sqrt{2}$, while infinite number of measurements in $\sigma_x$ and $\sigma_y$ Bloch plane may arrive at an even less upper bound, $\eta\leq 2/\pi$; for three-dimensional case, joint measurements on the three mutually unbiased bases (MUB) $\sigma_x$, $\sigma_y$, and $\sigma_z$ give $\eta\leq 1/\sqrt{3}$, while infinite number of measurements in the $MUB$ Bloch space may give $\eta\leq 1/2$, see the Appendix for details. These are the best results so far.

For dimension-three Werner and isotropic states, the parametrization may write
\begin{align}
\rho_{\mathrm{W}} & = \frac{1-\eta}{9}\mathds{1} \otimes \mathds{1} + \frac{\eta}{3}\sum_{ i\neq j}^3 |\psi_{ij}^-\rangle \langle\psi_{ij}^-|\; , \\
\rho_{\mathrm{ISO}} & = \frac{1-\eta}{9} \mathds{1} \otimes \mathds{1}+ \eta |\psi^{+}\rangle \langle \psi^+| \; .
\end{align}
Here $|\psi^-_{ij}\rangle = \frac{1}{\sqrt{2}}(|ij\rangle -|ji\rangle)$ and  $|\psi^+\rangle = \frac{1}{\sqrt{3}} \sum_{i=1}^N |ii\rangle$. For Werner state the number of degrees of freedom of $3\times 3$ observables, equals the number of SU$(3$) generators, is much higher than the number of measurement in 3-dimensional MUB. And hence from Theorem \ref{Theorem-steering} the MUB measurement result would be trivial. For isotropic state, the latest research based on entropic uncertainty relation predicts the steering inequality $\eta > 1/2$ \cite{Steering-Guhne-pra}, while QMS calculation shows that the steerability will exhibits at $\eta > \frac{3\sqrt{5}+1}{16} \sim 0.4818$. Results of the steerability for two- and three-dimensional Werner and isotropic states are summarized in Table I.
\begin{center}\centerline{{\bf Table 1:} The QMS predicted values of $\eta$ for non-steerable states.}
\begin{tabular}{|c|c|c|c|}
  \hline
\multirow{2}{*}{\diagbox{States}{Measurements} } &  \multicolumn{2}{|c|}{$N=2$}   & $N=3$ \\ \cline{2-3}  \cline{4-4}
 & 2D & 3D & MUB
 \\ \hline
\rule{0pt}{15pt}  $\rho_{\mathrm{W}}$  &  \multirow{2}{*}{$\displaystyle \eta \leq   \frac{2}{\pi}$} &  \multirow{2}{*}{$\displaystyle \eta \leq \frac{1}{2}$}   &  $\eta \leq 1$ \\ \cline{1-1}  \cline{4-4}
\rule{0pt}{15pt}   $\rho_{\mathrm{ISO}}$ &    &    & $\eta \leq \frac{3\sqrt{5}+1}{16}$  \\
  \hline
\end{tabular}
\end{center}

The above results reveal that the non-local character of steerability relative predominantly to the degrees of freedom of measured observables, but rather simply the number of observables.

\subsection{The separability in QMS}

By definition, a state $\rho_{AB}$ is separable if and only if it can be decomposed as
\begin{align}
\rho_{AB} = \sum_{\lambda} \kappa_{\lambda}\cdot \rho^{(\lambda)} \otimes \sigma^{(\lambda)} \; ,
\end{align}
given coefficients $\kappa_{\lambda}\ge 0$. Here $\rho^{(\lambda)}$ and $\sigma^{(\lambda)}$ are density matrices. For measurements $X$ and $X'$ on $Alice$ and $Y$ and $Y'$ on $Bob$, the quantum magic square description goes as
\begin{align}
m_{i_1i_2j_1j_2} & = \sum_{\lambda} \kappa_{\lambda} \cdot \left[\tilde{p}_{i_1}^{(\lambda)}(x) \tilde{p}_{i_2}^{(\lambda)}(x')\right] \left[\tilde{q}_{j_1}^{(\lambda)}(y) \tilde{q}_{j_2}^{(\lambda)}(y')\right]  =\sum_{\lambda} \kappa_{\lambda} \cdot m^{(\lambda)}_{(i_1i_2)(j_1j_2)}\; . \label{Sep-jjkk}
\end{align}
Here, the distribution vectors with tildes satisfy the uncertainty relations  \cite{Maj-opt}
\begin{align}
\vec{\tilde{p}}^{\,(\lambda)}(x) \oplus \vec{\tilde{p}}^{\,(\lambda)}(x') \prec \vec{s}(x,x') \; , \; \vec{\tilde{q}}^{\,(\lambda)}(y) \oplus \vec{\tilde{q}}^{\,(\lambda)}(y') \prec \vec{s}(y,y') \; . \label{S-element-maj}
\end{align}
It is interesting to compare the difference between two-way non-steering state and the separable state. The former in QMS scheme writes
\begin{align}
m_{i_1i_2j_1j_2} & = \sum_{\lambda} \kappa_{\lambda} \cdot \left[p^{(\lambda)}_{i_1}(x) p^{(\lambda)}_{i_2}(x')\right] \left[\tilde{q}^{(\lambda)}_{j_1}(y) \tilde{q}^{(\lambda)}_{j_2}(y')\right] =\sum_{\lambda} \kappa_{\lambda} \cdot m_{i_1i_2(j_1j_2)}^{(\lambda)} \; , \label{t-w-steering1} \\
m_{i_1i_2j_1j_2} & = \sum_{\lambda} \kappa_{\lambda} \cdot \left[\tilde{p}^{(\lambda)}_{i_1}(x) \tilde{p}^{(\lambda)}_{i_2}(x')\right] \left[q^{(\lambda)}_{j_1}(y) q^{(\lambda)}_{j_2}(y')\right]  = \sum_{\lambda} \kappa_{\lambda} \cdot m_{(i_1i_2)j_1j_2}^{(\lambda)} \; , \label{t-w-steering2}
\end{align}
where the tilde term is constrained by the uncertainty relation (\ref{S-element-maj}). It is now evident that the difference between the separability and two-way non-steering shows up in the difference of (\ref{t-w-steering1})-(\ref{t-w-steering2}) with (\ref{Sep-jjkk}). For separable state, there remains two uncertainty relations for each $m_{(i_1i_2)(j_1j_2)}^{(\lambda)}$, whereas it needs only one for the two-way non-steering case, no matter the decomposition in term of $m_{i_1i_2(j_1j_2)}^{(\lambda)}$ or $m_{(i_1i_2)j_1j_2}^{(\lambda)}$.

\section{Summary}

In this work we proposed a novel scheme, the quantum magic square, to characterize the quantum entanglement in form of high dimensional probability tensor, whose marginal distributions can reproduce all the desired joint measurements on certain quantum state. The tensor shows different inner structures as per the strength of quantum non-locality, from which the QMS helps us to get the tangible effects. To distinguish Bell local state from Bell non-local one, we may construct the QMS for any possible observables. In this scheme, the uncertainty relations between certain tensor components distinguish non-steering state from the Bell local state. The difference between the separable state with the non-steering state lies in having more constraints on tensor components in the form of uncertainty relation.

The QMS scheme has the merit of employing each individual component of the distribution vectors in form of the direct sum majorization uncertainty relation, while in other scalar-function-based detection criteria in the measurement of quantum entanglement, the variance, entropy, or other scalar measures, the quantumness of the state is usually averaged. In this sense, QMS sets up a general framework for the study of the quantum non-locality, including separability, non-separability, two/one-way non-steering, steering, Bell locality, and Bell non-locality, etc. This work is just a pioneer study on this scheme, further investigations on its application, for instance whether or not the quantum contextuality can still vindicate its effectiveness, are expected.

\section*{Acknowledgements}
\noindent
This work was supported in part by the Ministry of Science and Technology of the Peoples' Republic of China(2015CB856703); by the Strategic Priority Research Program of the Chinese Academy of Sciences, Grant No.XDB23030100; and by the National Natural Science Foundation of China(NSFC) under the Grants 11975236 and 11635009.

\newpage
\setcounter{figure}{0}
\renewcommand{\thefigure}{S\arabic{figure}}
\setcounter{equation}{0}
\renewcommand\theequation{S\arabic{equation}}
\setcounter{theorem}{0}
\renewcommand{\thetheorem}{S\arabic{theorem}}
\setcounter{observation}{0}
\renewcommand{\theobservation}{S\arabic{observation}}
\setcounter{proposition}{0}
\renewcommand{\theproposition}{S\arabic{proposition}}
\setcounter{lemma}{0}
\renewcommand{\thelemma}{S\arabic{lemma}}
\setcounter{corollary}{0}
\renewcommand{\thecorollary}{S\arabic{corollary}}
\setcounter{section}{0}
\renewcommand{\thesection}{S\arabic{section}}

\appendix{\bf \LARGE Appendix}

We present the detailed proofs for demonstration of various non-localities via quantum magic square.

\section{Bell non-locality and the magic squares}

\subsection{The CHSH inequality from the magic square}

\begin{figure}\centering
\scalebox{0.7}{\includegraphics{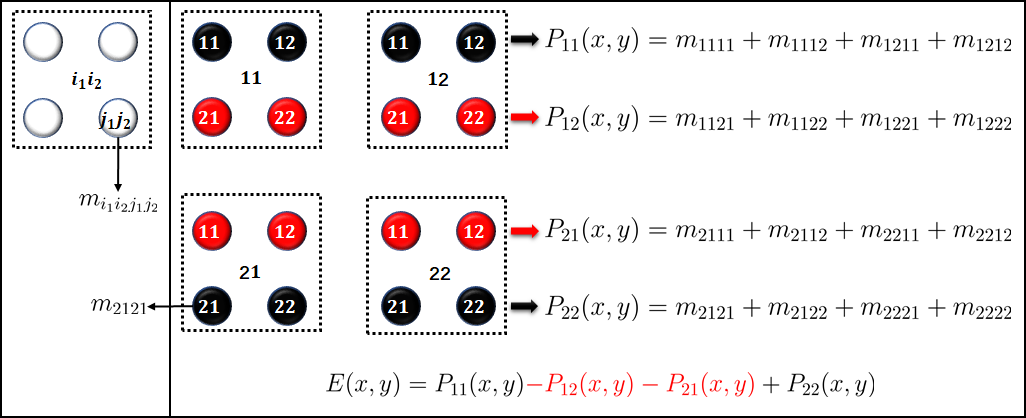}}
\caption{{\bf The correlation function $E(x,y)$ in magic squares.} Here each node in the dashed squares represents a component of the probability tensor $m_{i_1i_2j_1j_2}$. The black nodes are summed in the calculation of the correlation $E(x,y)$, while the red ones are subtracted.} \label{S-Exy-nodes}
\end{figure}

For the bipartite qubit states, the correlation functions for joint observables $X$ and $Y$ on particles $A$ and $B$ reads
\begin{align}
E(x,y) := P_{11}(x,y) - P_{12}(x,y) - P_{21}(x,y) + P_{22}(x,y) \; .
\end{align}
The CHSH inequality is the constraint by the local realistic theory on the four correlation functions
\begin{align}
\left| E(x,y) - E(x,y') + E(x',y) + E(x',y') \right| \leq 2 \; .
\end{align}
In the magic square scheme, the joint distributions can be obtained form the constructed probability tensor
\begin{align}
m_{i_1i_2j_1j_2} = \sum_{\lambda} \kappa_{\lambda} \cdot \left[ p^{(\lambda)}_{i_1}(x) p^{(\lambda)}_{i_2}(x') \right] \left[ q^{(\lambda)}_{j_1}(y) q^{(\lambda)}_{j_2}(y') \right] \; .
\end{align}
All the joint distributions can be derived from the above tensor,
\begin{align}
P_{i_1j_1}(x,y) = \sum_{i_2,j_2} m_{i_1i_2j_1j_2} \; , \; P_{i_1j_2}(x,y') = \sum_{i_2,j_1} m_{i_1i_2j_1j_2} \; , \\
P_{i_2j_1}(x',y) = \sum_{i_1,j_2} m_{i_1i_2j_1j_2} \; , \; P_{i_2j_2}(x',y') = \sum_{i_1,j_1} m_{i_1i_2j_1j_2} \; .
\end{align}
The correlation functions then can be built from the tensor $m_{i_1i_2j_1j_2}$, see Figure \ref{S-Exy-nodes}.

\begin{figure}\centering
\scalebox{0.7}{\includegraphics{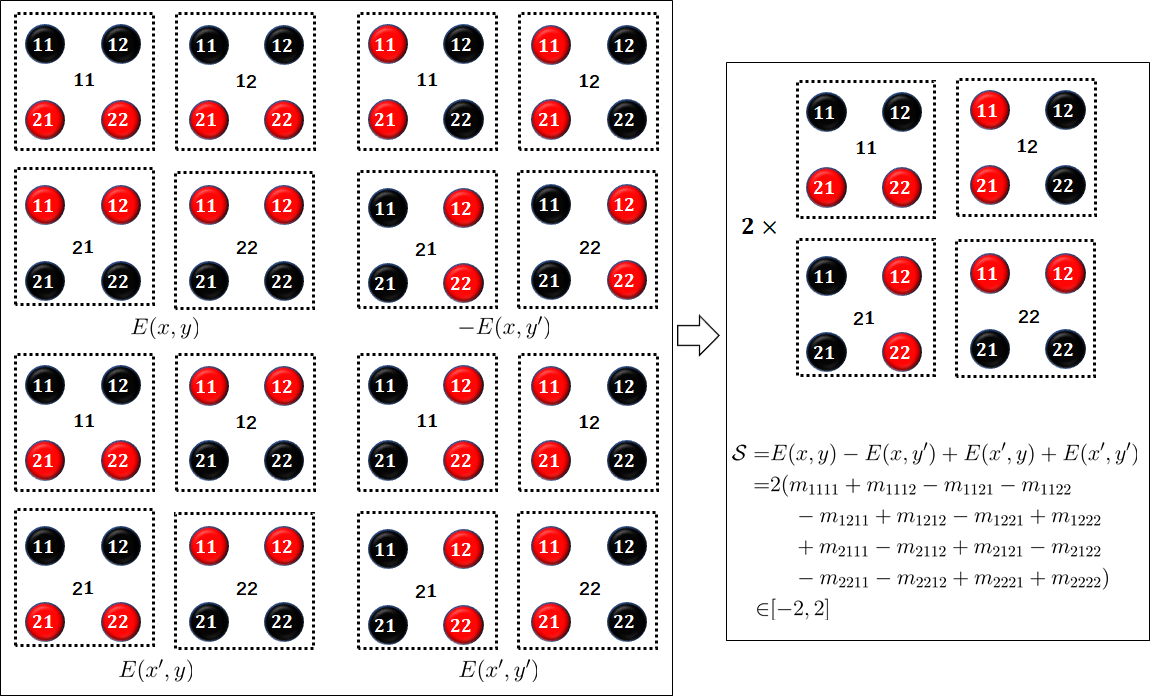}}
\caption{{\bf The CHSH inequality via magic squares.} In the magic square representation of the correlation functions, the CHSH inequality can be obtained by summation over the nodes.} \label{S-CHSH-nodes}
\end{figure}

All the four correlation functions can be obtained, see Figure \ref{S-CHSH-nodes}. Within the magic squares, see Figure \ref{S-CHSH-nodes}, the CHSH inequality now becomes,
\begin{align}
\mathcal{S} = & E(x,y) -E(x,y') + E(x',y) +E(x',y')  \nonumber \\
=& 2[ (m_{1111}+m_{1112} -m_{1121}-m_{1122})+( -m_{1211}+m_{1212}-m_{1221}+m_{1222}) \nonumber \\
& \hspace{0.5cm} +(m_{2111}-m_{2112}+m_{2121}-m_{2122})+ (-m_{2211}-m_{2212}+m_{2221}+ m_{2222}) ] \nonumber \\
=& 2 \left( \sum_{\{\mathrm{black}\}}m_{i_1i_2j_1j_2} - \sum_{\{\mathrm{red}\}} m_{i_1i_2j_1j_2} \right)
\in [-2,2]  \;.
\end{align}
Let $a = \sum_{\{\mathrm{black}\}}m_{i_1i_2j_1j_2}$, $b=\sum_{\{\mathrm{red}\}} m_{i_1i_2j_1j_2}$, in the last line of the above equation we have used the following fact
\begin{align}
a \in [0,1] \; ,\; b\in [0,1]\; ,\; a+b = 1 \; \Rightarrow \; 2(a-b) \in [-2,2] \; .
\end{align}
The local realism are described by the existence of probability tensor, the magic square, $m_{i_1i_2j_1j_2}$. Thus the contradiction between the quantum theory and local realism is captured by our scheme.

\subsection{GHZ theorem from the magic squares}

\begin{figure}\centering
\scalebox{0.5}{\includegraphics{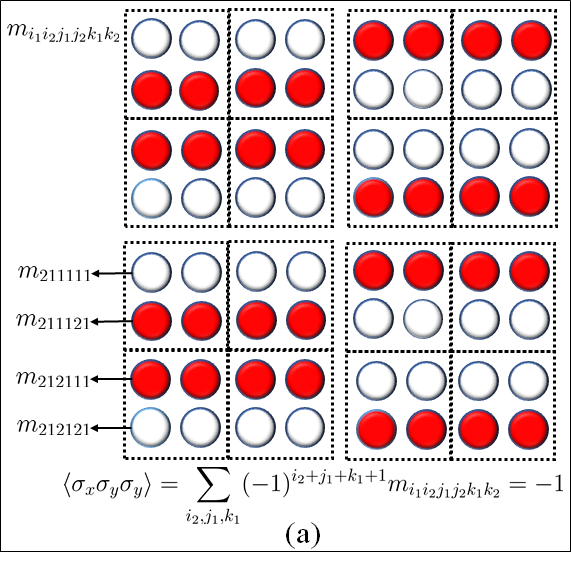}}
\scalebox{0.5}{\includegraphics{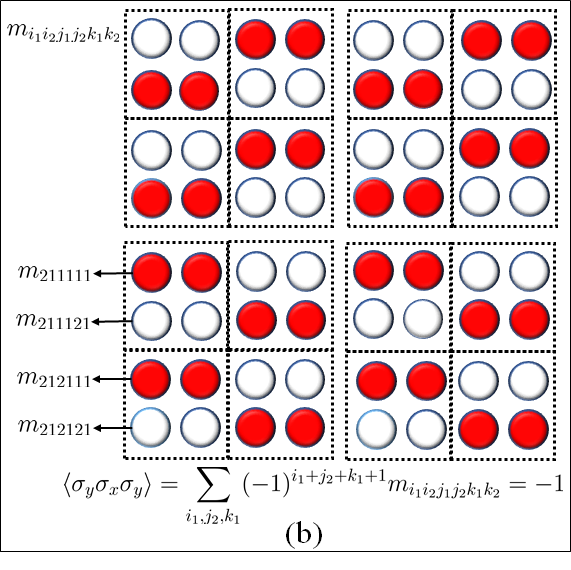}}
\scalebox{0.5}{\includegraphics{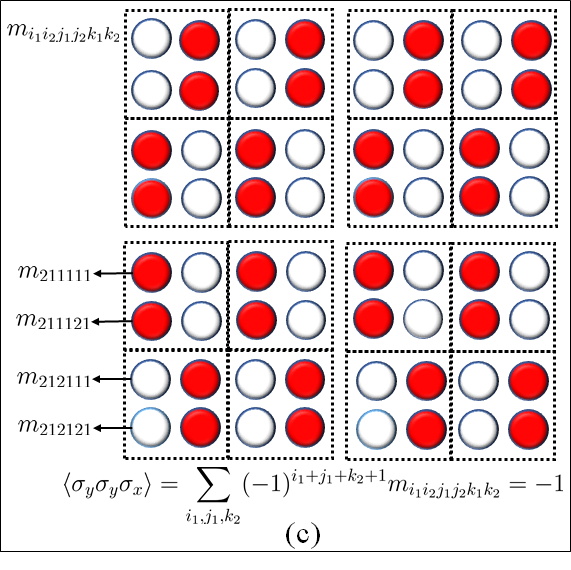}} \\
\scalebox{0.5}{\includegraphics{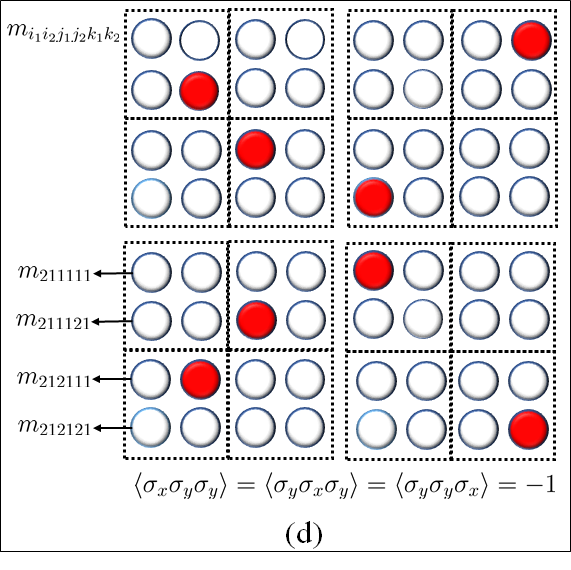}}
\scalebox{0.5}{\includegraphics{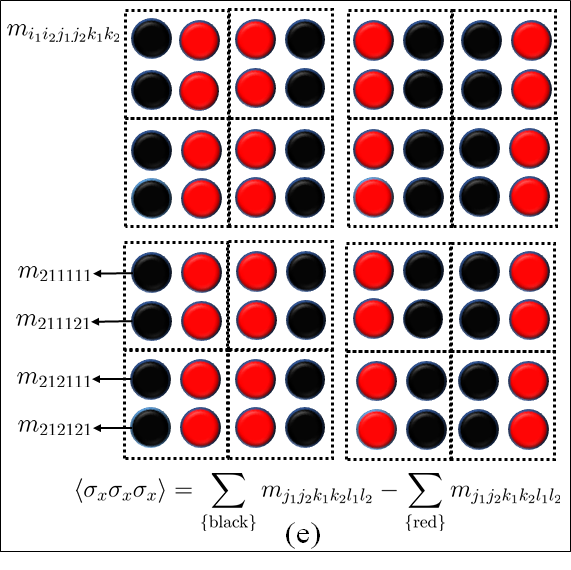}}
\caption{{\bf The GHZ theorem with magic squares.} The expectation values for $\langle \sigma_x\sigma_y\sigma_y\rangle = -1$, $\langle \sigma_y\sigma_x\sigma_y\rangle=-1$, and $\langle \sigma_y\sigma_y\sigma_x\rangle  =-1$ are expressed as the sum of the red nodes in (a)-(c) respectively, where the blank nodes are zeros. In (d), the expectation value for $\langle \sigma_x\sigma_y\sigma_y \rangle = \langle \sigma_y\sigma_x\sigma_y \rangle = \langle \sigma_y\sigma_y\sigma_x
\rangle = -1$ must be expressed as the sum of the red nodes that appears in (a)-(c) simultaneously. And the red nodes in (d) can only predict $\langle \sigma_x\sigma_x\sigma_x\rangle = -1$ in (e). } \label{S-GHZ-node}
\end{figure}

Given the observables $X=\sigma_x$ and $Y=\sigma_y$ on each site, we may construct the following magic square
\begin{align}
m_{i_1i_2j_1j_2k_1k_2} = \sum_{\lambda} \kappa_{\lambda} \cdot \left[ p_{i_1}^{(\lambda)}(x) p_{i_2}^{(\lambda)}(y)\right] \left[ q_{j_1}^{(\lambda)}(x) q_{j_2}^{(\lambda)}(y)\right] \left[ r_{k_1}^{(\lambda)}(x) r_{k_2}^{(\lambda)}(y)\right] \; .
\end{align}
Taking the joint observation $\sigma_x\sigma_y\sigma_y$ as example, the expectation value can be evaluated as
\begin{align}
\langle \sigma_x\sigma_y\sigma_y \rangle = \sum_{i_2,j_1,k_1} (-1)^{i_2+j_1+k_1+1} m_{i_1i_2j_1j_2k_1k_2} \;. \label{S-eq-xyy}
\end{align}
Here we adopt the convention that for the eigenvalues of $\pm 1$, we have $p_{i_1=+}^{(\lambda)}(x) + p^{(\lambda)}_{i_1=-}(x) =1$ and $\langle \sigma_x\rangle = \sum_{\lambda} \kappa_{\lambda} \cdot [p^{(\lambda)}_{+}(x)+(-1)p^{(\lambda)}_{-}(x)]$. Because $m_{i_1i_2j_1j_2k_1k_2}$ are positive semidefinite, equation (\ref{S-eq-xyy}) can be reexpressed as, see Figure \ref{S-GHZ-node}(a),
\begin{align}
\langle \sigma_x\sigma_y\sigma_y \rangle = \sum_{\{\mathrm{blank}\}} m_{i_1i_2j_1j_2k_1k_2} -  \sum_{\{\mathrm{red}\}} m_{i_1i_2j_1j_2k_1k_2} \; .
\end{align}
If $\langle \sigma_x\sigma_y\sigma_y \rangle = -1$ then the blank nodes of $m_{i_1i_2j_1j_2k_1k_2}$ are zeros, and
\begin{align}
\langle \sigma_x\sigma_y\sigma_y \rangle =  -  \sum_{\{\mathrm{red}\}} m_{i_1i_2j_1j_2k_1k_2} = -1\; .
\end{align}
This can be shown via the following: Let $a=\sum_{\{\mathrm{blank}\}} m_{i_1i_2j_1j_2k_1k_2}$ and $b = \sum_{\{\mathrm{red}\}} m_{i_1i_2j_1j_2k_1k_2}$, then from $a-b=-1$ and $1\geq a,b \geq 0$, we have $a=0$ and $b=1$, see Figure \ref{S-GHZ-node}(a).

The expectation values of $\langle \sigma_y\sigma_x\sigma_y \rangle$ and $\langle \sigma_y\sigma_y\sigma_x \rangle$ can be obtained similarly, see Figure \ref{S-GHZ-node}(b)-(c). However, for the GHZ state $|\psi\rangle = \frac{1}{\sqrt{2}}(|000\rangle - |111\rangle)$, all the three terms are $-1$, i.e., $\langle \sigma_x \sigma_{y} \sigma_y \rangle = \langle \sigma_y \sigma_{x} \sigma_y\rangle = \langle \sigma_y\sigma_{y} \sigma_x\rangle = - 1$. This puts a strong constraint on the magic square $m_{i_1i_2j_1j_2k_1k_2}$, where only a few nodes are nonzero, see Figure \ref{S-GHZ-node}(d). With the nonzero nodes for $\langle \sigma_x \sigma_{y} \sigma_y \rangle = \langle \sigma_y \sigma_{x} \sigma_y\rangle = \langle \sigma_y\sigma_{y} \sigma_x\rangle = - 1$, the expectation value for $\sigma_x\sigma_x\sigma_x$ can only be
\begin{align}
\langle \sigma_x\sigma_x\sigma_x \rangle = \sum_{i_2,j_2,k_2} (-1)^{i_2+j_2+k_2+1} m_{i_1i_2j_1j_2k_1k_2} = -1 \; ,
\end{align}
see Figure \ref{S-GHZ-node}(e). This contradicts the predictions of quantum mechanics.

\section{Quantum steering and the magic squares}

\subsection{The closed path summation in magic squares}

\begin{figure}\centering
\scalebox{0.6}{\includegraphics{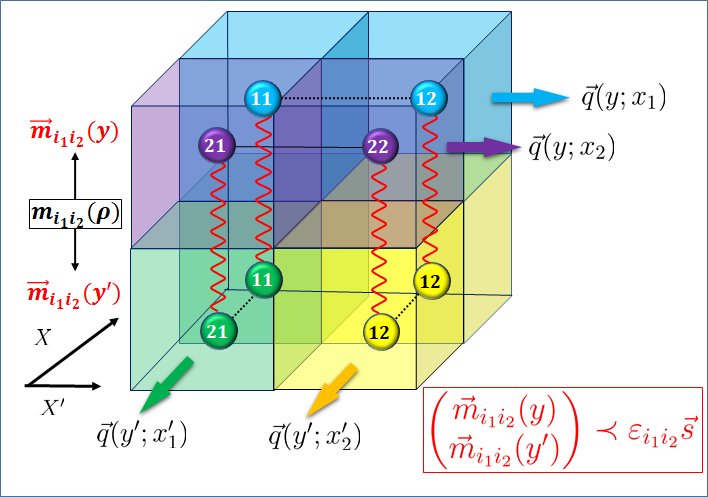}}
\caption{ } \label{S-qubit-S}
\end{figure}

For bipartite qubit systems, if the state is Bell local, then there exists the magic square description for the system. Further if $A$ cannot steer the system $B$, then the there exist a local hidden state (LHS) description of the system. For two measurements $X$ and $X'$ on $A$'s side, the magic square for the LHS is
\begin{align}
m_{i_1i_2}(\rho) = \sum_{\lambda} \kappa_{\lambda} \cdot \left[ p_{i_1}^{(\lambda)}(x) p_{i_2}^{(\lambda)}(x')\right] \rho^{(\lambda)} = \sum_{\lambda} \kappa_{\lambda} \cdot m_{i_1i_2}^{(\lambda)}(\rho)\; .
\end{align}
Here $m_{i_1i_2}(\rho)$ is a superposition of $m_{i_1i_2}^{(\lambda)}(\rho) = p^{(\lambda)}_{i_1}(x) p_{i_2}^{(\lambda)}(x')\rho^{(\lambda)}$ and is an unnormalized state represented by cubic cells in Figure \ref{S-qubit-S}. In the Figure \ref{S-qubit-S}, two layers of magic squares are piled up for two measurements $Y$ and $Y'$ on $B$'s side. There exists the following majorization uncertainty relation for each $m^{(\lambda)}_{i_1i_2}(\rho)$
\begin{align}
\vec{m}^{(\lambda)}_{i_1i_2}(y) \oplus \vec{m}^{(\lambda)}_{i_1i_2}(y') =  \varepsilon^{(\lambda)}_{i_1i_2}  D^{(\lambda)}\vec{s} \, \prec \varepsilon^{(\lambda)}_{i_1i_2}  \vec{s} \; . \label{S-Maj-cell}
\end{align}
where $\varepsilon^{(\lambda)}_{i_1i_2} = p^{(\lambda)}_{i_1}(x) p_{i_2}^{(\lambda)}(x')$ and $D^{(\lambda)}$ is a doubly stochastic matrix. Here we are legitimate to rearrange the components of the distributions $\vec{m}^{(\lambda)}_{i_1i_2}(y)$ and $\vec{m}^{(\lambda)}_{i_1i_2}(y)$ in descending order respectively.

\begin{figure}\centering
\scalebox{0.7}{\includegraphics{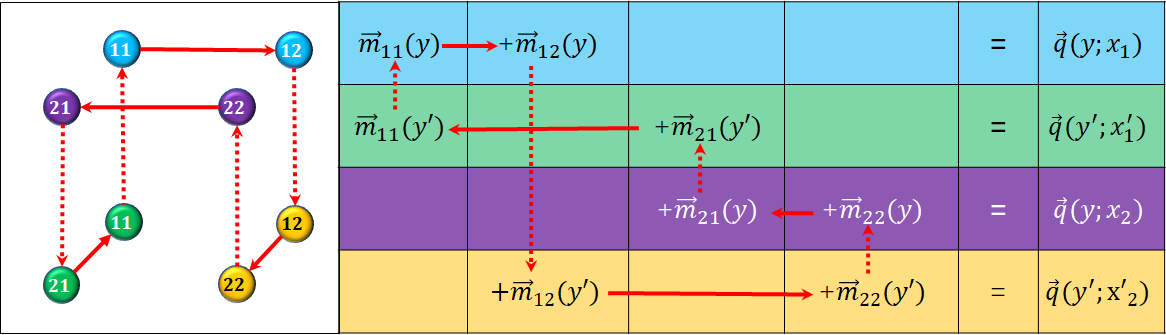}}
\caption{{\bf The closed path summation.} The closed path summation over the nodes of the magic squares gives the observable effects that is originally hidden in the uncertainty relation between the nodes. The observable effects can be used to distinguish the non-steering from Bell locality. } \label{S-qubit-close-path}
\end{figure}

In order to get a physical observable effects from the uncertainty relation imposed on each tensor elements, we make summations along a closed path as shown in Figure \ref{S-qubit-close-path}. Simplification of the closed path summation gives
\begin{align}
\begin{pmatrix}
\vec{q}(y;x_1)\\
\vec{q}(y';x_1') \\
\vec{q}(y;x_2) \\
\vec{q}(y';x_2')
\end{pmatrix} = \begin{pmatrix}
\vec{m}_{11}(y) \\
\vec{m}_{11}(y') \\
\vec{m}_{22}(y) \\
\vec{m}_{22}(y')
\end{pmatrix} +
\begin{pmatrix}
\vec{m}_{12}(y) \\
\vec{m}_{21}(y') \\
\vec{m}_{21}(y) \\
\vec{m}_{12}(y')
\end{pmatrix} \; .
\end{align}
Or equivalently
\begin{align}
\sum_{\lambda} \kappa_{\lambda} \begin{pmatrix}
\vec{q}^{\,(\lambda)}(y;x_1)\\
\vec{q}^{\,(\lambda)}(y';x_1') \\
\vec{q}^{\,(\lambda)}(y;x_2) \\
\vec{q}^{\,(\lambda)}(y';x_2')
\end{pmatrix} = \sum_{\lambda} \kappa_{\lambda} \left[\begin{pmatrix}
\vec{m}^{(\lambda)}_{11}(y) \\
\vec{m}^{(\lambda)}_{11}(y') \\
\vec{m}^{(\lambda)}_{22}(y) \\
\vec{m}^{(\lambda)}_{22}(y')
\end{pmatrix} +
\begin{pmatrix}
\vec{m}^{(\lambda)}_{12}(y) \\
\vec{m}^{(\lambda)}_{21}(y') \\
\vec{m}^{(\lambda)}_{21}(y) \\
\vec{m}^{(\lambda)}_{12}(y')
\end{pmatrix} \right] \; . \label{S-two-qubit-close}
\end{align}
For each $\lambda$ in equation (\ref{S-two-qubit-close}) we have
\begin{align}
\begin{pmatrix}
\vec{q}^{\,(\lambda)}(y;x_1)\\
\vec{q}^{\,(\lambda)}(y';x_1') \\
\vec{q}^{\,(\lambda)}(y;x_2) \\
\vec{q}^{\,(\lambda)}(y';x_2')
\end{pmatrix}  = & \left[ \begin{pmatrix}
\varepsilon^{(\lambda)}_{11} & 0 & 0 & 0 \\
0 & \varepsilon^{(\lambda)}_{11}  & 0 & 0 \\
0 & 0 & \varepsilon^{(\lambda)}_{22}  & 0 \\
0 & 0 & 0 & \varepsilon^{(\lambda)}_{22}
\end{pmatrix}
\begin{pmatrix}
D^{(\lambda)}_{11} & D^{(\lambda)}_{12} & 0 & 0 \\
D^{(\lambda)}_{21} & D^{(\lambda)}_{22} & 0 & 0 \\
0 & 0 & D^{(\lambda)}_{11} & D^{(\lambda)}_{12} \\
0 & 0 & D^{(\lambda)}_{21} & D^{(\lambda)}_{22}
\end{pmatrix} \right. \nonumber \\
& \left. + \begin{pmatrix}
\varepsilon^{(\lambda)}_{12} & 0 & 0 & 0 \\
0 & \varepsilon^{(\lambda)}_{21}  & 0 & 0 \\
0 & 0 & \varepsilon^{(\lambda)}_{21}  & 0 \\
0 & 0 & 0 & \varepsilon^{(\lambda)}_{12}
\end{pmatrix}
\begin{pmatrix}
D^{(\lambda)}_{11} & D^{(\lambda)}_{12} & 0 & 0 \\
0 & 0 & D^{(\lambda)}_{21} & D^{(\lambda)}_{22} \\
0 & 0 & D^{(\lambda)}_{11} & D^{(\lambda)}_{12} \\
D^{(\lambda)}_{21} & D^{(\lambda)}_{22} & 0 & 0
\end{pmatrix} \right]
\begin{pmatrix}
\vec{s} \\
\vec{s}
\end{pmatrix}  \nonumber \\  =  & \left[ \begin{pmatrix}
\varepsilon^{(\lambda)}_{11}* & \varepsilon^{(\lambda)}_{11}* & 0 & 0 \\
\varepsilon^{(\lambda)}_{11}* & \varepsilon^{(\lambda)}_{11}* & 0 & 0 \\
0 & 0 & \varepsilon^{(\lambda)}_{22} * & \varepsilon^{(\lambda)}_{22}* \\
0 & 0 & \varepsilon^{(\lambda)}_{22}* & \varepsilon^{(\lambda)}_{22}*
\end{pmatrix}   +
\begin{pmatrix}
\varepsilon^{(\lambda)}_{12}* & \varepsilon^{(\lambda)}_{12}* & 0 & 0 \\
0 & 0 & \varepsilon^{(\lambda)}_{21}* & \varepsilon^{(\lambda)}_{21}* \\
0 & 0 & \varepsilon^{(\lambda)}_{21}* & \varepsilon^{(\lambda)}_{21}* \\
\varepsilon^{(\lambda)}_{12}* & \varepsilon^{(\lambda)}_{12}* & 0 & 0
\end{pmatrix} \right]
\begin{pmatrix}
\vec{s} \\
\vec{s}
\end{pmatrix}  \;. \label{S-qubit-example}
\end{align}
Here $D_{ij}^{(\lambda)}$ are submatrices of $D^{(\lambda)}$ with dimensions $2\times 2$. Define $\vec{e} \equiv (1,1)^{\mathrm{T}}$, we notice that
\begin{align}
\begin{pmatrix}
\vec{s} \\
\vec{s}
\end{pmatrix} = \vec{e} \otimes \vec{s} =
(\mathds{E}\otimes \mathds{1}) (\vec{e} \otimes \vec{s}) = \mathds{E}\otimes \mathds{1}
\begin{pmatrix}
\vec{s} \\
\vec{s}
\end{pmatrix} \; ,
\end{align}
where $\mathds{E}= \frac{1}{2}\begin{pmatrix}
1 & 1 \\
1 & 1
\end{pmatrix}$. Multiply $\mathds{E}\otimes \mathds{1}$ from the left to the equation (\ref{S-qubit-example}), we have
\begin{align}
\mathds{E}\otimes \mathds{1} \begin{pmatrix}
\vec{q}^{\,(\lambda)}(y;x_1)\\
\vec{q}^{\,(\lambda)}(y';x_1') \\
\vec{q}^{\,(\lambda)}(y;x_2) \\
\vec{q}^{\,(\lambda)}(y';x_2')
\end{pmatrix} = & \left[ \mathds{E}\otimes \mathds{1} \begin{pmatrix}
\varepsilon^{(\lambda)}_{11}* & \varepsilon^{(\lambda)}_{11}* & 0 & 0 \\
\varepsilon^{(\lambda)}_{11}* & \varepsilon^{(\lambda)}_{11}* & 0 & 0 \\
0 & 0 & \varepsilon^{(\lambda)}_{22} * & \varepsilon^{(\lambda)}_{22}* \\
0 & 0 & \varepsilon^{(\lambda)}_{22}* & \varepsilon^{(\lambda)}_{22}*
\end{pmatrix} \mathds{E}\otimes \mathds{1} \right. \nonumber \\ & \left. +
\mathds{E}\otimes \mathds{1}\begin{pmatrix}
\varepsilon^{(\lambda)}_{12}* & \varepsilon^{(\lambda)}_{12}* & 0 & 0 \\
0 & 0 & \varepsilon^{(\lambda)}_{21}* & \varepsilon^{(\lambda)}_{21}* \\
0 & 0 & \varepsilon^{(\lambda)}_{21}* & \varepsilon^{(\lambda)}_{21}* \\
\varepsilon^{(\lambda)}_{12}* & \varepsilon^{(\lambda)}_{12}* & 0 & 0
\end{pmatrix} \mathds{E}\otimes \mathds{1} \right] \begin{pmatrix}
\vec{s} \\
\vec{s}
\end{pmatrix} \nonumber \\
= &  \left[\frac{\varepsilon^{(\lambda)}_{11} + \varepsilon^{(\lambda)}_{22}}{2}
\begin{pmatrix}
\frac{1}{2}D^{(\lambda)} & \frac{1}{2}D^{(\lambda)} \\
\frac{1}{2}D^{(\lambda)} & \frac{1}{2}D^{(\lambda)}
\end{pmatrix} +  \frac{\varepsilon^{(\lambda)}_{12} + \varepsilon^{(\lambda)}_{21}}{2}
\begin{pmatrix}
\frac{1}{2}D^{(\lambda)} & \frac{1}{2}D^{(\lambda)} \\
\frac{1}{2}D^{(\lambda)} & \frac{1}{2}D^{(\lambda)}
\end{pmatrix}  \right]
\begin{pmatrix}
\vec{s} \\
\vec{s}
\end{pmatrix} \nonumber \\
= & \frac{1}{2} \mathcal{D}^{(\lambda)}
\begin{pmatrix}
\vec{s} \\
\vec{s}
\end{pmatrix} \; . \label{S-two-qubit-trick2}
\end{align}
Here $\mathcal{D}^{(\lambda)} := \begin{pmatrix}
\frac{1}{2}D^{(\lambda)} & \frac{1}{2}D^{(\lambda)} \\
\frac{1}{2}D^{(\lambda)} & \frac{1}{2}D^{(\lambda)}
\end{pmatrix}$ is also a doubly stochastic matrix and what we actually get from equation (\ref{S-two-qubit-trick2}) is
\begin{align}
\begin{pmatrix}
\vec{q}^{\,(\lambda)}(y;x_1) + \vec{q}^{\,(\lambda)}(y;x_2)\\
\vec{q}^{\,(\lambda)}(y';x_1') + \vec{q}^{\,(\lambda)}(y';x_2')\\
\end{pmatrix} & =
\begin{pmatrix}
\vec{q}^{\,(\lambda)}(y) p^{(\lambda)}_1(x) + \vec{q}^{\,(\lambda)}(y) p^{(\lambda)}_2(x)\\
\vec{q}^{\,(\lambda)}(y') p^{(\lambda)}_1(x') + \vec{q}^{\,(\lambda)}(y') p^{(\lambda)}_2(x')\\
\end{pmatrix} \nonumber \\ & = \begin{pmatrix}
\vec{q}^{\,(\lambda)}(y) \\
\vec{q}^{\,(\lambda)}(y') \\
\end{pmatrix}
= D^{(\lambda)} \vec{s} \; . \label{S-two-qubit-stochastic}
\end{align}
All the $\vec{q}^{\,(\lambda)}(y,y')$ can be rearranged in descending order respectively, i.e.,
\begin{align}
\begin{pmatrix}
\vec{q}^{\,(\lambda)\downarrow}(y) p^{(\lambda)}_1(x) + \vec{q}^{\,(\lambda)\downarrow}(y) p^{(\lambda)}_2(x)\\
\vec{q}^{\,(\lambda)\downarrow}(y') p^{(\lambda)}_1(x') + \vec{q}^{\,(\lambda)\downarrow}(y') p^{(\lambda)}_2(x')\\
\end{pmatrix}=D^{(\lambda)} \vec{s}\; .
\end{align}
(Here the doubly stochastic matrices $D^{(\lambda)}$ are the same as that of equation (\ref{S-two-qubit-stochastic}) up to permutations of rows.) There exist the following properties for majorization (see chapter 6 of Ref.\cite{S-Majorization-Book})
\begin{align}
\sum_{\lambda} \kappa_{\lambda} \cdot \vec{q}^{\,(\lambda)}(y) p^{(\lambda)}_{i_1}(x) & \prec \sum_{\lambda} \kappa_{\lambda} \cdot \vec{q}^{\,(\lambda)\downarrow}(y) p^{(\lambda)}_{i_1}(x) \; ,\\
\sum_{\lambda} \kappa_{\lambda} \cdot \vec{q}^{\,(\lambda)}(y') p^{(\lambda)}_{i_2}(x') & \prec \sum_{\lambda} \kappa_{\lambda} \cdot \vec{q}^{\,(\lambda)\downarrow}(y') p^{(\lambda)}_{i_2}(x')\; ,
\end{align}
from which we can directly write
\begin{align}
\vec{q}^{\,\downarrow}(y;x_{i_1}) & = \left[\sum_{\lambda} \kappa_{\lambda} \cdot \vec{q}^{\,(\lambda)}(y) p^{(\lambda)}_{i_1}(x)\right]^{\,\downarrow} \prec \sum_{\lambda} \kappa_{\lambda} \cdot \vec{q}^{\,(\lambda)\downarrow}(y) p^{(\lambda)}_{i_1}(x) \; , \\
\vec{q}^{\,\downarrow}(y';x'_{i_2}) & = \left[\sum_{\lambda} \kappa_{\lambda} \cdot \vec{q}^{\,(\lambda)}(y') p^{(\lambda)}_{i_2}(x')\right]^{\,\downarrow} \prec \sum_{\lambda} \kappa_{\lambda} \cdot \vec{q}^{\,(\lambda)\downarrow}(y') p^{(\lambda)}_{i_2}(x') \;.
\end{align}
Again, using the summation properties for the majorization, we get the conditional majorization uncertainty relation for the LHS model
\begin{align}
\begin{pmatrix}
\sum_{i_1} \vec{q}^{\,\downarrow}(y;x_{i_1}) \\
\sum_{i_2} \vec{q}^{\,\downarrow}(y';x'_{i_2})
\end{pmatrix} \prec \vec{s}\;. \label{S-quNit-two-sum}
\end{align}
Here for the qubit state we have
\begin{align}
\begin{pmatrix}
\vec{q}^{\,\downarrow}(y;x_1)+ \vec{q}^{\,\downarrow}(y;x_2) \\
\vec{q}^{\,\downarrow}(y';x_1') + \vec{q}^{\,\downarrow}(y';x_2')
\end{pmatrix} =
\begin{pmatrix}
\vec{q}^{\,\downarrow}(y|x_1) p(x_1) + \vec{q}^{\,\downarrow}(y|x_2) p(x_2)\\
\vec{q}^{\,\downarrow}(y'|x_1') p(x_1') + \vec{q}^{\,\downarrow}(y'|x_2') p(x_2')
\end{pmatrix} \prec \vec{s} \; , \label{S-qubit-prec}
\end{align}
where $\vec{s}$ is the least upper bound of the direct sum majorization uncertainty relation for $Y$ and $Y'$ \cite{S-Maj-Un-Relation}.

We consider the two qubit Werner state as example
\begin{align}
\rho_{\mathrm{W}} = \frac{1-\eta}{4} \mathds{1} \otimes \mathds{1} + \eta|\psi^-_{12}\rangle \langle \psi^-_{12}| \; ,
\end{align}
where $|\psi^-_{12}\rangle = \frac{1}{\sqrt{2}}(|12\rangle - |21\rangle)$. For joint measurements $X =\sigma_x$ and $X'=\sigma_y$ on $A$'s side and $Y=\sigma_x$ and $Y'=\sigma_y$ on $B$'s side, equation (\ref{S-qubit-prec}) becomes
\begin{align}
(\frac{1+\eta}{2}, \frac{1-\eta}{2})\oplus (\frac{1+\eta}{2}, \frac{1-\eta}{2})
\prec (1,\frac{\sqrt{2}}{2},\frac{2-\sqrt{2}}{2},0) \; .
\end{align}
Here $\vec{s}$ is the least upper bound for the uncertainty relation of $\sigma_x$ and $\sigma_y$ \cite{S-Maj-Un-Relation}.
From the majorization  uncertainty relation, the sum of the first four largest terms on the left and right hand sides must satisfy
\begin{align}
2\times \frac{1+\eta}{ 2} \leq 1+\frac{\sqrt{2}}{2} = 2\cos^2\frac{\pi}{8} \; \Rightarrow \eta \leq \frac{\sqrt{2}}{2}\; \;. \label{Werner-qubit-xy}
\end{align}
The advantage of our method is that a large number of measurements can be performed on both sides. For example, we may perform measurements with interval of $\mathrm{d}\theta$ along the two dimensional Bloch vector space of $\sigma_x$ and $\sigma_y$,
\begin{align}
\frac{\pi}{\mathrm{d}\theta} \times \frac{1+\eta}{2} \leq \sum_{i=1}^{\frac{\pi}{\mathrm{d}\theta}} \cos^2\frac{\theta_i}{2} \; ,
\end{align}
where equation (\ref{Werner-qubit-xy}) corresponds to the case of $\theta_i\in\{\pm \frac{\pi}{4}\}$ and the interval $\mathrm{d}\theta = \pi/2$. In the limiting case of $\mathrm{d}\theta \to 0$ we have
\begin{align}
\frac{1+\eta}{2} & \leq \frac{1}{\pi} \lim_{\mathrm{d}\theta \to 0} \sum_{i=1}^{\frac{\pi}{\mathrm{d}\theta}} \cos^{2}\frac{\theta}{2} \,\mathrm{d}\theta \nonumber \\
& = \frac{1}{\pi} \int_{-\frac{\pi}{2}}^{\frac{\pi}{2}} \cos^{2}\frac{\theta}{2} \, \mathrm{d}\theta = \frac{\pi+2}{2\pi}\; \Rightarrow \; \eta\leq \frac{2}{\pi}\;.
\end{align}
This is the best value that we can have for measurements along the plane of the Bloch vector space spanned by $\sigma_x$ and $\sigma_y$.

\subsection{High dimensional observables}

For arbitrary $N$-dimensional measurements $X$ and $X'$, the two joint measurements $X$-$Y$ and $X'$-$Y'$ on the bipartite state have the following
\begin{align}
\left(
\displaystyle \sum_{i_1=1}^N \vec{q}^{\,\downarrow}(y|x_{i_1}) p(x_{i_1})\right) \oplus
\left(\displaystyle \sum_{i_2=1}^N \vec{q}^{\,\downarrow}(y'|x_{i_2}') p(x_{i_2}')
\right) \prec \vec{s} \; , \label{S-bi-3}
\end{align}
where $\vec{s}$ is the least upper bound of the majorization uncertainty relation for observables $Y$ and $Y'$. Equation (\ref{S-bi-3}) can be readily seen from equation (\ref{S-quNit-two-sum}).

\subsection{Multiple observables}

\subsubsection{Three qubit observables}

\begin{figure}\centering
\scalebox{0.7}{\includegraphics{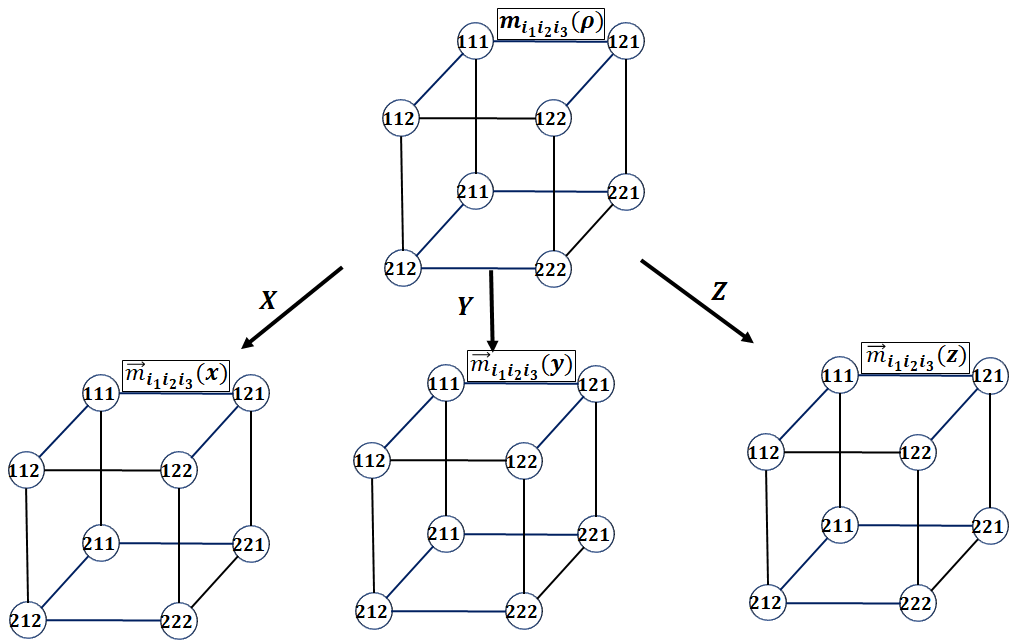}}
\caption{{\bf Magic squares for non-steering in bipartite qubit system.} Three different measurements $X$, $Y$, $Z$ may be performed on $B$'s side, and the corresponding nodes $m_{i_1i_2i_3}(\rho)$ become the distribution vectors $\vec{m}_{i_1i_2i_3}(x)$, $\vec{m}_{i_1i_2i_3}(y)$, or $\vec{m}_{i_1i_2i_3}(z)$.} \label{S-bi-xyz}
\end{figure}

\begin{figure}\centering
\scalebox{0.75}{\includegraphics{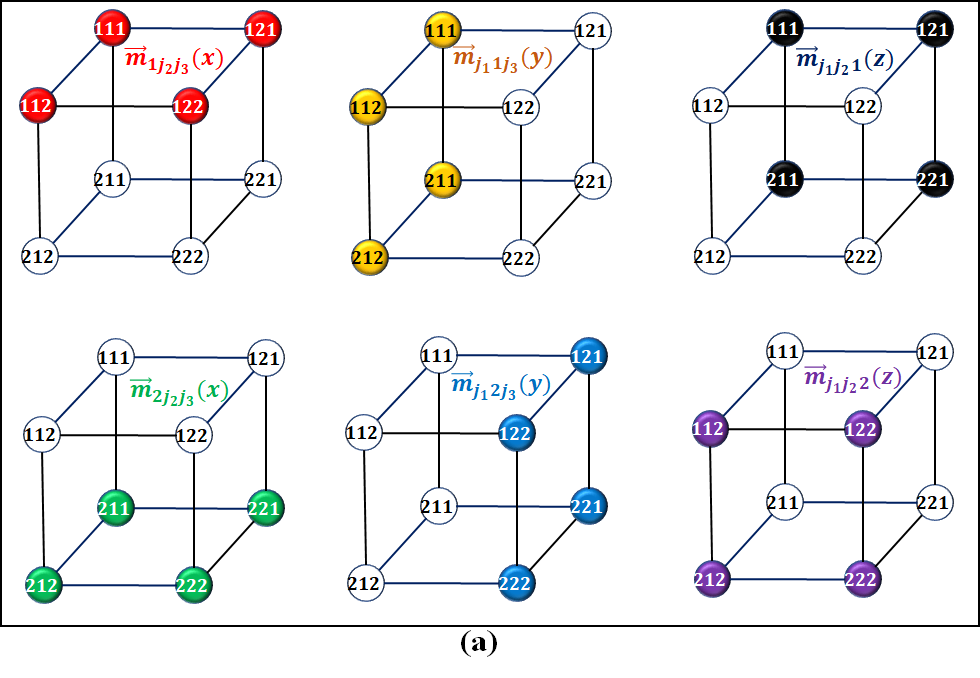}} \\ \vspace{0.5cm}
\scalebox{0.62}{\includegraphics{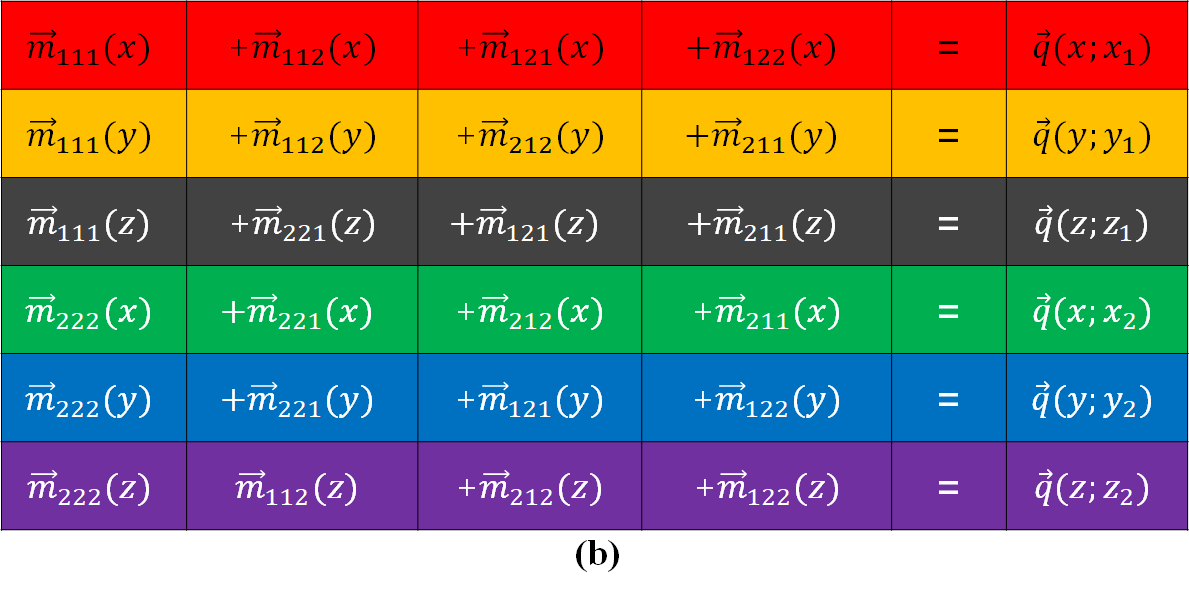}}
\caption{{\bf Magic squares for quantum steering with three observables $X$, $Y$, and $Z$ on each side.} {\bf (a)} The magic squares for conditional distributions; {\bf (b)} The closed path summation for the conditional distributions.} \label{S-XYZ-steering-MC}
\end{figure}

Here we consider the three dichotomic observables, $X=\sigma_x$, $Y=\sigma_y$, and $Z=\sigma_z$, in bipartite qubit system. The non-steering condition from $A$ to $B$ can be expressed in terms of magic squares, see Figure \ref{S-bi-xyz}
\begin{align}
m_{i_1i_2i_3}(\rho) = \sum_{\lambda} \kappa_{\lambda} \cdot \left[ p^{(\lambda)}_{i_1}(x) p^{(\lambda)}_{i_2}(y) p^{(\lambda)}_{i_3}(z) \right] \rho^{(\lambda)} = \sum_{\lambda} \kappa_{\lambda} \cdot m_{i_1i_2i_3}^{(\lambda)}(\rho)\; .
\end{align}
Three different measurements $X$, $Y$, $Z$ may be performed on $B$'s side, and the corresponding nodes $m_{i_1i_2i_3}(\rho)$ may become the distribution vectors $\vec{m}_{i_1i_2i_3}(x)$, $\vec{m}_{i_1i_2i_3}(y)$, or $\vec{m}_{i_1i_2i_3}(z)$, see Figure \ref{S-bi-xyz}.
From Figure \ref{S-bi-xyz} it is clear that the distributions $\vec{m}_{i_1i_2i_3}(x,y,z)$ satisfy the following uncertainty relation
\begin{align}
\begin{pmatrix}
\vec{m}^{(\lambda)}_{i_1i_2i_3}(x)  \\
\vec{m}^{(\lambda)}_{i_1i_2i_3}(y) \\
\vec{m}^{(\lambda)}_{i_1i_2i_3}(z)
\end{pmatrix}= \varepsilon^{(\lambda)}_{i_1i_2i_3} D^{(\lambda)} \vec{s} \prec \varepsilon^{(\lambda)}_{i_1i_2i_3} \vec{s}  \; ,
\end{align}
where $\varepsilon^{(\lambda)}_{i_1i_2i_3} =p^{(\lambda)}_{i_1}(x) p^{(\lambda)}_{i_2}(y) p^{(\lambda)}_{i_3}(z)$ and $D^{(\lambda)}$ is a doubly stochastic matrix of dimensions $6\times 6$. $\vec{s}$ is the least upper bound of the direct sum majorization relation for $X$, $Y$, and $Z$ whose value can be found in \cite{S-Maj-Un-Relation}. In Figure \ref{S-XYZ-steering-MC}(b), the first column may be expressed as
\begin{align}
& \left(
\vec{m}^{(\lambda)}_{111}(x) ,
\vec{m}^{(\lambda)}_{111}(y) ,
\vec{m}^{(\lambda)}_{111}(z) ,
\vec{m}^{(\lambda)}_{222}(x) ,
\vec{m}^{(\lambda)}_{222}(y) ,
\vec{m}^{(\lambda)}_{222}(z)
\right)^{\mathrm{T}} \nonumber \\ =&
\begin{pmatrix}
\varepsilon^{(\lambda)}_{111}D^{(\lambda)} & 0 \\
0 & \varepsilon^{(\lambda)}_{222} D^{(\lambda)}
\end{pmatrix}
\begin{pmatrix}
\vec{s} \\ \vec{s}
\end{pmatrix} \; .
\end{align}
Similar expresses exist for the next three columns. And the equation
\begin{align}
\begin{pmatrix}
\vec{q}^{\,(\lambda)}(x;x_1) \\
\vec{q}^{\,(\lambda)}(y;y_1) \\
\vec{q}^{\,(\lambda)}(z;z_1) \\
\vec{q}^{\,(\lambda)}(x;x_2) \\
\vec{q}^{\,(\lambda)}(y;y_2) \\
\vec{q}^{\,(\lambda)}(z;z_2)
\end{pmatrix} :=  
\begin{pmatrix}
\vec{m}_{111}^{(\lambda)}(x) \\
\vec{m}_{111}^{(\lambda)}(y) \\
\vec{m}_{111}^{(\lambda)}(z) \\
\vec{m}_{222}^{(\lambda)}(x) \\
\vec{m}_{222}^{(\lambda)}(y) \\
\vec{m}_{222}^{(\lambda)}(z)
\end{pmatrix} + \begin{pmatrix}
\vec{m}_{112}^{(\lambda)}(x) \\
\vec{m}_{112}^{(\lambda)}(y) \\
\vec{m}_{221}^{(\lambda)}(z) \\
\vec{m}_{221}^{(\lambda)}(x) \\
\vec{m}_{221}^{(\lambda)}(y) \\
\vec{m}_{112}^{(\lambda)}(z)
\end{pmatrix} + \begin{pmatrix}
\vec{m}_{121}^{(\lambda)}(x) \\
\vec{m}_{212}^{(\lambda)}(y) \\
\vec{m}_{121}^{(\lambda)}(z) \\
\vec{m}_{212}^{(\lambda)}(x) \\
\vec{m}_{121}^{(\lambda)}(y) \\
\vec{m}_{212}^{(\lambda)}(z)
\end{pmatrix} + 
\begin{pmatrix}
\vec{m}_{122}^{(\lambda)}(x) \\
\vec{m}_{211}^{(\lambda)}(y) \\
\vec{m}_{211}^{(\lambda)}(z) \\
\vec{m}_{211}^{(\lambda)}(x) \\
\vec{m}_{122}^{(\lambda)}(y) \\
\vec{m}_{122}^{(\lambda)}(z)
\end{pmatrix} \;,
\end{align} 
which give the Figure \ref{S-XYZ-steering-MC}(b) when summed over $\lambda$, becomes
\begin{align}
& \left(
\vec{q}^{\,(\lambda)}(x;x_1) ,
\vec{q}^{\,(\lambda)}(y;y_1) ,
\vec{q}^{\,(\lambda)}(z;z_1) ,
\vec{q}^{\,(\lambda)}(x;x_2) ,
\vec{q}^{\,(\lambda)}(y;y_2) ,
\vec{q}^{\,(\lambda)}(z;z_2)
\right)^{\mathrm{T}} \nonumber \\ = &
\left[
\begin{pmatrix}
\varepsilon^{(\lambda)}_{111} & 0 & 0 & 0 & 0 & 0 \\
0 & \varepsilon^{(\lambda)}_{111} & 0 & 0 & 0 & 0 \\
0 & 0 & \varepsilon^{(\lambda)}_{111} & 0 & 0 & 0 \\
0 & 0 & 0 & \varepsilon^{(\lambda)}_{222} & 0 & 0 \\
0 & 0 & 0 & 0 & \varepsilon^{(\lambda)}_{222} & 0 \\
0 & 0 & 0 & 0 & 0 & \varepsilon^{(\lambda)}_{222}
\end{pmatrix}
\begin{pmatrix}
D^{(\lambda)}_{11} & D^{(\lambda)}_{12} & D^{(\lambda)}_{13} & 0 & 0 & 0 \\
D^{(\lambda)}_{21} & D^{(\lambda)}_{22} & D^{(\lambda)}_{23} & 0 & 0 & 0 \\
D^{(\lambda)}_{31} & D^{(\lambda)}_{32} & D^{(\lambda)}_{33} & 0 & 0 & 0 \\
0 & 0 & 0 & D^{(\lambda)}_{11} & D^{(\lambda)}_{12} & D^{(\lambda)}_{13} \\
0 & 0 & 0 & D^{(\lambda)}_{21} & D^{(\lambda)}_{22} & D^{(\lambda)}_{23} \\
0 & 0 & 0 & D^{(\lambda)}_{31} & D^{(\lambda)}_{32} & D^{(\lambda)}_{33}
\end{pmatrix} \right. \nonumber \\ & + \begin{pmatrix}
\varepsilon^{(\lambda)}_{112} & 0 & 0 & 0 & 0 & 0 \\
0 & \varepsilon^{(\lambda)}_{112} & 0 & 0 & 0 & 0 \\
0 & 0 & \varepsilon^{(\lambda)}_{221} & 0 & 0 & 0 \\
0 & 0 & 0 & \varepsilon^{(\lambda)}_{221} & 0 & 0 \\
0 & 0 & 0 & 0 & \varepsilon^{(\lambda)}_{221} & 0 \\
0 & 0 & 0 & 0 & 0 & \varepsilon^{(\lambda)}_{112}
\end{pmatrix}
\begin{pmatrix}
D^{(\lambda)}_{11} & D^{(\lambda)}_{12} & D^{(\lambda)}_{13} & 0 & 0 & 0 \\
D^{(\lambda)}_{21} & D^{(\lambda)}_{22} & D^{(\lambda)}_{23} & 0 & 0 & 0 \\
0 & 0 & 0 & D^{(\lambda)}_{31} & D^{(\lambda)}_{32} & D^{(\lambda)}_{33} \\
0 & 0 & 0 & D^{(\lambda)}_{11} & D^{(\lambda)}_{12} & D^{(\lambda)}_{13} \\
0 & 0 & 0 & D^{(\lambda)}_{21} & D^{(\lambda)}_{22} & D^{(\lambda)}_{23} \\
D^{(\lambda)}_{31} & D^{(\lambda)}_{32} & D^{(\lambda)}_{33} & 0 & 0 & 0
\end{pmatrix}  \nonumber \\
& + \begin{pmatrix}
\varepsilon^{(\lambda)}_{121} & 0 & 0 & 0 & 0 & 0 \\
0 & \varepsilon^{(\lambda)}_{212} & 0 & 0 & 0 & 0 \\
0 & 0 & \varepsilon^{(\lambda)}_{121} & 0 & 0 & 0 \\
0 & 0 & 0 & \varepsilon^{(\lambda)}_{212} & 0 & 0 \\
0 & 0 & 0 & 0 & \varepsilon^{(\lambda)}_{121} & 0 \\
0 & 0 & 0 & 0 & 0 & \varepsilon^{(\lambda)}_{212}
\end{pmatrix}
\begin{pmatrix}
D^{(\lambda)}_{11} & D^{(\lambda)}_{12} & D^{(\lambda)}_{13} & 0 & 0 & 0 \\
0 & 0 & 0 & D^{(\lambda)}_{21} & D^{(\lambda)}_{22} & D^{(\lambda)}_{23} \\
D^{(\lambda)}_{31} & D^{(\lambda)}_{32} & D^{(\lambda)}_{33} & 0 & 0 & 0 \\
0 & 0 & 0 & D^{(\lambda)}_{11} & D^{(\lambda)}_{12} & D^{(\lambda)}_{13} \\
D^{(\lambda)}_{21} & D^{(\lambda)}_{22} & D^{(\lambda)}_{23} &0 & 0 & 0 \\
0 & 0 & 0 & D^{(\lambda)}_{31} & D^{(\lambda)}_{32} & D^{(\lambda)}_{33}
\end{pmatrix} \nonumber \\
& \left. +  \begin{pmatrix}
\varepsilon^{(\lambda)}_{122} & 0 & 0 & 0 & 0 & 0 \\
0 & \varepsilon^{(\lambda)}_{211} & 0 & 0 & 0 & 0 \\
0 & 0 & \varepsilon^{(\lambda)}_{211} & 0 & 0 & 0 \\
0 & 0 & 0 & \varepsilon^{(\lambda)}_{211} & 0 & 0 \\
0 & 0 & 0 & 0 & \varepsilon^{(\lambda)}_{122} & 0 \\
0 & 0 & 0 & 0 & 0 & \varepsilon^{(\lambda)}_{122}
\end{pmatrix}
\begin{pmatrix}
D^{(\lambda)}_{11} & D^{(\lambda)}_{12} & D^{(\lambda)}_{13} & 0 & 0 & 0 \\
0 & 0 & 0 & D^{(\lambda)}_{21} & D^{(\lambda)}_{22} & D^{(\lambda)}_{23}  \\
0 & 0 & 0 & D^{(\lambda)}_{31} & D^{(\lambda)}_{32} & D^{(\lambda)}_{23}  \\
0 & 0 & 0 & D^{(\lambda)}_{11} & D^{(\lambda)}_{12} & D^{(\lambda)}_{13} \\
D^{(\lambda)}_{21} & D^{(\lambda)}_{22} & D^{(\lambda)}_{23} &0 & 0 & 0 \\
D^{(\lambda)}_{31} & D^{(\lambda)}_{32} & D^{(\lambda)}_{33} & 0 & 0 & 0
\end{pmatrix} \right]
\begin{pmatrix}
\vec{s} \\
\vec{s}
\end{pmatrix} \; . \label{S-tri-ob-exp}
\end{align}
The same trick of multiplying $\mathds{E}\otimes \mathds{1}$ in the qubit case applies equally well to equation (\ref{S-tri-ob-exp}) with $\mathds{1}$ having the dimensions of $3\times 3$, and we have
\begin{align}
& \mathds{E}\otimes \mathds{1} \left(
\vec{q}^{\,(\lambda)}(x;x_1) ,
\vec{q}^{\,(\lambda)}(y;y_1) ,
\vec{q}^{\,(\lambda)}(z;z_1) ,
\vec{q}^{\,(\lambda)}(x;x_2) ,
\vec{q}^{\,(\lambda)}(y;y_2) ,
\vec{q}^{\,(\lambda)}(z;z_2)
\right)^{\mathrm{T}} \nonumber \\
= &
\left[ \frac{\varepsilon^{(\lambda)}_{111}+ \varepsilon^{(\lambda)}_{222}}{2}
\begin{pmatrix}
\frac{D^{(\lambda)}}{2} &  \frac{D^{(\lambda)}}{2} \\
\frac{D^{(\lambda)}}{2} & \frac{D^{(\lambda)}}{2}
\end{pmatrix} + \frac{\varepsilon^{(\lambda)}_{112}+ \varepsilon^{(\lambda)}_{221}}{2}
\begin{pmatrix}
\frac{D^{(\lambda)}}{2} &  \frac{D^{(\lambda)}}{2} \\
\frac{D^{(\lambda)}}{2} & \frac{D^{(\lambda)}}{2}
\end{pmatrix} \right. \nonumber \\
& \left. + \frac{\varepsilon^{(\lambda)}_{121}+ \varepsilon^{(\lambda)}_{212}}{2}
\begin{pmatrix}
\frac{D^{(\lambda)}}{2} &  \frac{D^{(\lambda)}}{2} \\
\frac{D^{(\lambda)}}{2} & \frac{D^{(\lambda)}}{2}
\end{pmatrix}  + \frac{\varepsilon^{(\lambda)}_{122}+ \varepsilon^{(\lambda)}_{211}}{2}
\begin{pmatrix}
\frac{D^{(\lambda)}}{2} &  \frac{D^{(\lambda)}}{2} \\
\frac{D^{(\lambda)}}{2} & \frac{D^{(\lambda)}}{2}
\end{pmatrix} \right]
\begin{pmatrix}
\vec{s} \\
\vec{s}
\end{pmatrix} \; ,
\end{align}
which is just
\begin{align}
\begin{pmatrix}
\vec{q}^{\,(\lambda)\downarrow}(x;x_1) + \vec{q}^{\,(\lambda)\downarrow}(x;x_2)\\
\vec{q}^{\,(\lambda)\downarrow}(y;y_1) + \vec{q}^{\,(\lambda)\downarrow}(y;y_2)\\
\vec{q}^{\,(\lambda)\downarrow}(z;z_1) + \vec{q}^{\,(\lambda)\downarrow}(z;z_2)
\end{pmatrix} = D^{\lambda} \vec{s}
\end{align}
Similarly, we have the conditional majorization uncertainty relation for three observables
\begin{align}
\begin{pmatrix}
\vec{q}^{\,\downarrow}(x|x_1)p(x_1) + \vec{q}^{\,\downarrow}(x|x_2)p(x_2) \\
\vec{q}^{\,\downarrow}(y|y_1)p(y_1) + \vec{q}^{\,\downarrow}(y|y_2)p(y_2) \\
\vec{q}^{\,\downarrow}(z|z_1)p(z_1) + \vec{q}^{\,\downarrow}(z|z_2)p(z_2)
\end{pmatrix} \prec   \vec{s} \; .  \label{S-bi-3-ori}
\end{align}
where $\vec{s}$ is the least upper bound for the majorization uncertainty relation of $X$, $Y$, and $Z$ \cite{S-Maj-Un-Relation}.

We consider the two qubit Werner state,
\begin{align}
\rho_{\mathrm{W}} = \frac{1-\eta}{4} \mathds{1} \otimes \mathds{1} + \eta |\psi^-\rangle \langle \psi^-| =
\begin{pmatrix}
\frac{1-\eta}{4} & 0 & 0 & 0 \\
0 & \frac{1+\eta}{4} & \frac{-\eta}{2} & 0 \\
0 & \frac{-\eta}{2} & \frac{1+\eta}{4} & 0 \\
0 & 0 & 0 & \frac{1-\eta}{4}
\end{pmatrix} \; .
\end{align}
Because the Werner state is symmetric $\rho_{\mathrm{W}} = (u\otimes u) \rho_{\mathrm{W}} (u^{\dag}\otimes u)$, so we have \begin{align}
\vec{q}^{\downarrow}(x;x_i) = \begin{pmatrix}
\frac{1+\eta}{4} \\
\frac{1-\eta}{4}
\end{pmatrix} \; ,\; \vec{q}^{\downarrow}(y;y_i) = \begin{pmatrix}
\frac{1+\eta}{4} \\
\frac{1-\eta}{4}
\end{pmatrix} \; , \; \vec{q}^{\downarrow}(z;z_i) = \begin{pmatrix}
\frac{1+\eta}{4} \\
\frac{1-\eta}{4}
\end{pmatrix}\; .
\end{align}
For the mutually unbiased basis of $X=\sigma_x$, $Y=\sigma_y$, and $Z=\sigma_{z}$, the steering criterion of equation (\ref{S-bi-3-ori}) now can be expressed as
\begin{align}
\begin{pmatrix}
\frac{1+\eta}{2} \\
\frac{1-\eta}{2}
\end{pmatrix} \oplus \begin{pmatrix}
\frac{1+\eta}{2} \\
\frac{1-\eta}{2}
\end{pmatrix} \oplus \begin{pmatrix}
\frac{1+\eta}{2} \\
\frac{1-\eta}{2}
\end{pmatrix} \prec  \vec{s} \; . \label{S-MUB-qubit}
\end{align}
Here $\vec{s}=(1, \frac{\sqrt{2}}{2}, \frac{1+\sqrt{3}-\sqrt{2}}{2}, \frac{1-\sqrt{3}+\sqrt{2}}{2},\frac{2-\sqrt{2}}{2},0)$ (see \cite{S-Maj-Un-Relation}). Considering the first six terms on both sides of equation (\ref{S-MUB-qubit}), we have
\begin{align}
3 \times \frac{1+\eta}{2} \leq \frac{3+\sqrt{3}}{2} \Rightarrow \eta \leq \frac{\sqrt{3}}{3} \; .
\end{align}
This value for $\eta$ agrees with that of Ref. \cite{S-Ste-entr-guhne}, where the $\frac{1}{\sqrt{3}}$ was regarded as optimal for the uncertainty relation with mutually unbiased basis. In the next, we shall show that our method provides the optimal result. Considering a regular icosahedron whose 12 vertices are on the unit sphere of the three dimensional Bloch space for qubit. Six measurements can be performed by adopting their Bloch vectors of the Hermitian operator to be the six pairs of the vertices. The sum of the first six terms of $\vec{s}$ can be calculated
\begin{align}
\sum_{i=1}^6 s_i = 1 + 5*\cos^2\frac{\theta}{2} = 1+ 5\times \frac{3+\sqrt{5}}{5+\sqrt{5}} = \frac{4\sqrt{5}+6}{\sqrt{5}+1} \; .
\end{align}
where $\theta$ is the angle between two near vertices.
\begin{align}
6\times \frac{1+\eta}{2} \leq \frac{4\sqrt{5}+6}{\sqrt{5}+1} \; \Rightarrow \; \eta \leq \frac{\sqrt{5}+3}{3(\sqrt{5}+1)} \sim 0.5393 \; .
\end{align}
This value is better than $\frac{1}{\sqrt{3}} \sim 0.5773$ \cite{S-Steering-Nat-exp}. Our Theorem also alow us to consider the more general case, when a large number of measurements are performed. We can write
\begin{align}
& \frac{2\pi}{\mathrm{d}\Omega} \times \frac{1+\eta}{2} \leq \sum_{i=1}^{\frac{2\pi}{\mathrm{d}\Omega}} \cos^2\frac{\theta_i}{2} \nonumber\\ \Rightarrow & \frac{1+\eta}{2} \leq \frac{1}{2\pi} \int_{0}^{\frac{\pi}{2}}\int_{0}^{2\pi} \cos^2\frac{\theta}{2}\, \sin\theta \mathrm{d}\theta\mathrm{d}\phi = \frac{3}{4}\nonumber \\
\Rightarrow & \eta \leq \frac{1}{2} \; .
\end{align}
Here $\mathrm{d}\Omega = \sin\theta\mathrm{d}\theta\mathrm{d}\phi$. In the limiting case, we get the necessary and sufficient value $\eta>1/2$ for steering.

\subsubsection{Three dimensional Werner and isotropic states}

The three dimensional Werner and isotropic state are
\begin{align}
\rho_{\mathrm{W}} & = \frac{1-\eta}{9}\mathds{1} \otimes \mathds{1} + \frac{\eta}{3}\sum_{ i\neq j}^3 |\psi_{ij}^-\rangle \langle\psi_{ij}^-| \\
\rho_{\mathrm{ISO}} & = \frac{1-\eta}{9} \mathds{1} \otimes \mathds{1}+ \eta |\psi^{+}\rangle \langle \psi^+| \; .
\end{align}
Here $|\psi^-_{ij}\rangle = \frac{1}{\sqrt{2}}(|ij\rangle -|ji\rangle)$ and  $|\psi^+\rangle = \frac{1}{\sqrt{3}} \sum_{i=1}^N |ii\rangle$. In form of matrices we have
\begin{align}
\rho_{\mathrm{W}} & =
\begin{pmatrix}
\frac{1-\eta}{9} &0&0&0&0&0&0&0&0 \\
0&\frac{2+\eta}{18}&0&\frac{-\eta}{6}&0&0&0&0&0\\
0&0&\frac{2+\eta}{18}&0&0&0&\frac{-\eta}{6}&0&0\\
0&\frac{-\eta}{6}&0&\frac{2+\eta}{18}&0&0&0&0&0\\
0&0&0&0&\frac{1-\eta}{9}&0&0&0&0\\
0&0&0&0&0&\frac{2+\eta}{18}&0&\frac{-\eta}{6}&0\\
0&0&\frac{-\eta}{6}&0&0&0&\frac{2+\eta}{18}&0&0\\
0&0&0&0&0&\frac{-\eta}{6}&0&\frac{2+\eta}{18}&0\\
0&0&0&0&0&0&0&0&\frac{1-\eta}{9}
\end{pmatrix} \; ,\\
\rho_{\mathrm{ISO}} & =
\begin{pmatrix}
\frac{1+2\eta}{9} &0&0&0&\frac{\eta}{3}&0&0&0&\frac{\eta}{3} \\
0&\frac{1-\eta}{9}&0&0&0&0&0&0&0\\
0&0&\frac{1-\eta}{9}&0&0&0&0&0&0\\
0&0&0&\frac{1-\eta}{9}&0&0&0&0&0\\
\frac{\eta}{3}&0&0&0&\frac{1+2\eta}{9}&0&0&0&\frac{\eta}{3}\\
0&0&0&0&0&\frac{1-\eta}{9}&0&0&0\\
0&0&0&0&0&0&\frac{1-\eta}{9}&0&0\\
0&0&0&0&0&0&0&\frac{1-\eta}{9}&0\\
\frac{\eta}{3}&0&0&0&\frac{\eta}{3}&0&0&0&\frac{1+2\eta}{9}
\end{pmatrix}\; .
\end{align}
Here we choose the following MUB
\begin{align}
\begin{pmatrix}
1&0&0\\
0&1&0\\
0&0&1
\end{pmatrix}\;,\;
\frac{1}{\sqrt{3}}
\begin{pmatrix}
1& e^{-i\omega} & e^{i\omega}\\
1& e^{i\omega}  & e^{-i\omega}\\
1& 1            & 1
\end{pmatrix}\;,\\
\frac{1}{\sqrt{3}}
\begin{pmatrix}
e^{-i\omega} & e^{i\omega} & 1\\
e^{-i\omega} & 1           & e^{i\omega}\\
1           & 1            & 1
\end{pmatrix}\;, \;
\frac{1}{\sqrt{3}}
\begin{pmatrix}
e^{i\omega} & 1            & e^{-i\omega}\\
e^{i\omega} & e^{-i\omega} & 1\\
1           & 1            & 1
\end{pmatrix}\; . \label{S-MUB-3}
\end{align}
With this four basis, we can compute that
\begin{align}
\sum_{i=1}^4 s_{i}^{\downarrow} = \frac{3+\sqrt{5}}{2} \; , \; \sum_{i=1}^8 s_{i}^{\downarrow}  = 4\; .
\end{align}
Therefore for non-steering isotropic states, we shall have
\begin{align}
4\times \frac{1+2\eta}{3} \leq \frac{3+\sqrt{5}}{2} \Longrightarrow \eta \leq \frac{3\sqrt{5}+1}{16} \; .
\end{align}
And for non-steering Werner states, we shall have
\begin{align}
8 \times \frac{2+\eta}{6} \leq 4 \Longrightarrow \eta \leq 1\;.
\end{align}
It is interesting to observe that the steering condition for Werner state is trivial. Our scheme gives the following explanation: 1. The degree of freedom of the MUB in equation (\ref{S-MUB-3}) is 4 and is much smaller than the degree of freedom of the observables ($3\times 3$ Hermitian matrices); 2. The steering in this MUB is more sensitivity to isotropic state.

\end{document}